\documentclass[final]{aa}

\usepackage{makecell}
\usepackage{graphicx}
\usepackage{changes}
\usepackage{txfonts}
\usepackage{lscape}
\usepackage{natbib}
\usepackage{subcaption}
\DeclareMathOperator{\sech}{sech}

\begin{document}
\title{Improving metallicity estimates for very metal-poor stars in the Gaia DR3 GSP-Spec catalog}
\author{Tadafumi Matsuno\inst{1}
\and
Else Starkenburg\inst{1}
\and
Eduardo Balbinot\inst{1}
\and
Amina Helmi\inst{1}
}
\institute{
   Kapteyn Astronomical Institute, University of Groningen, Landleven 12, 9747 AD Groningen, The Netherlands\\
   \email{matsuno@astro.rug.nl}
}
\abstract
{
In the latest Gaia data release (DR3), the GSP-Spec module has provided stellar parameters and chemical abundances measured from the RVS spectra alone.
\added{
However, the GSP-Spec parameters, including metallicity, for very metal-poor stars (VMP stars; $[\mathrm{Fe/H}]<-2$) suffer from parameter degeneracy due to a lack of information in their spectra, and hence have large measurement uncertainty and systematic offset.
Furthermore, the recommended quality cuts filter out the majority ($\sim80\%$) of the VMP stars because some of them are confused with hot stars, which could be affected by the grid border effect, or with cool K and M-type giants, whose analysis can be challenging due to significant molecular absorptions. 
}
}
{
We aim to
\added{provide more precise} metallicity estimates for VMP stars analysed by the GSP-Spec module \added{by taking photometric information into account in the analysis and breaking the degeneracy}.
}
{
\added{We reanalyzed FGK type stars in the GSP-Spec catalog by computing the Ca triplet equivalent widths from the published set of GSP-Spec stellar parameters.}
\added{We compare these recovered equivalent widths with the values directly measured from public Gaia RVS spectra and investigate the precision of the recovered values and the parameter range in which the recovered values are reliable.}
We then convert the recovered equivalent widths to metallicities adopting photometric temperatures and surface gravities that we derive based on Gaia and 2MASS catalogs.
}
{
\added{
The recovered equivalent widths agree with the directly measured values with a scatter of $0.05\,\mathrm{dex}$ for the stars that pass the GSP-Spec quality cuts. 
Among stars that were recommended to filter out, we observed a similar scatter for FGK-type stars initially misidentified as hot stars.
Contrarily, we found a poorer agreement in general for stars that the GSP-Spec identified as cool K and M-type giants, although we can still define subsets that show reasonable agreement.
At the low metallicity end ($[\mathrm{Fe/H}]<-1.5$), our metallicity estimates have a typical uncertainty of $0.18\,\mathrm{dex}$, which is about half of the quoted GSP-Spec metallicity uncertainty at the same metallicity.
Our metallicities also show better agreement with the high-resolution literature values than the original GSP-Spec metallicities at low metallicity; the scatter in the comparison decreases from $0.36-0.46\,\mathrm{dex}$ to $0.17-0.29\,\mathrm{dex}$ for stars that satisfy the GSP-Spec quality cuts.
While the GSP-Spec metallicities show increasing scatter when misidentified ``hot'' stars and the subsets of the ``cool K and M-type giants'' are included (up to $1.06\,\mathrm{dex}$), we can now identify them as FGK-type stars and provide metallicities that show a small scatter in the comparisons (up to $0.34\,\mathrm{dex}$).
Thanks to the recovery of these originally misclassified stars, we increase the number of VMP stars with reliable metallicity by a factor of $2-3$.
}
}
{
The inclusion of photometric information greatly contributes to breaking parameter degeneracy, enabling precise metallicity estimates for VMP stars from Gaia RVS spectra. 
We produce a publicly-available catalog of bright metal-poor stars suitable for high-resolution follow-up.
}
\maketitle

\section{Introduction}
\begin{figure}
\centering
  \begin{subfigure}[t]{0.45\textwidth}
  \centering
  \includegraphics[width=\linewidth]{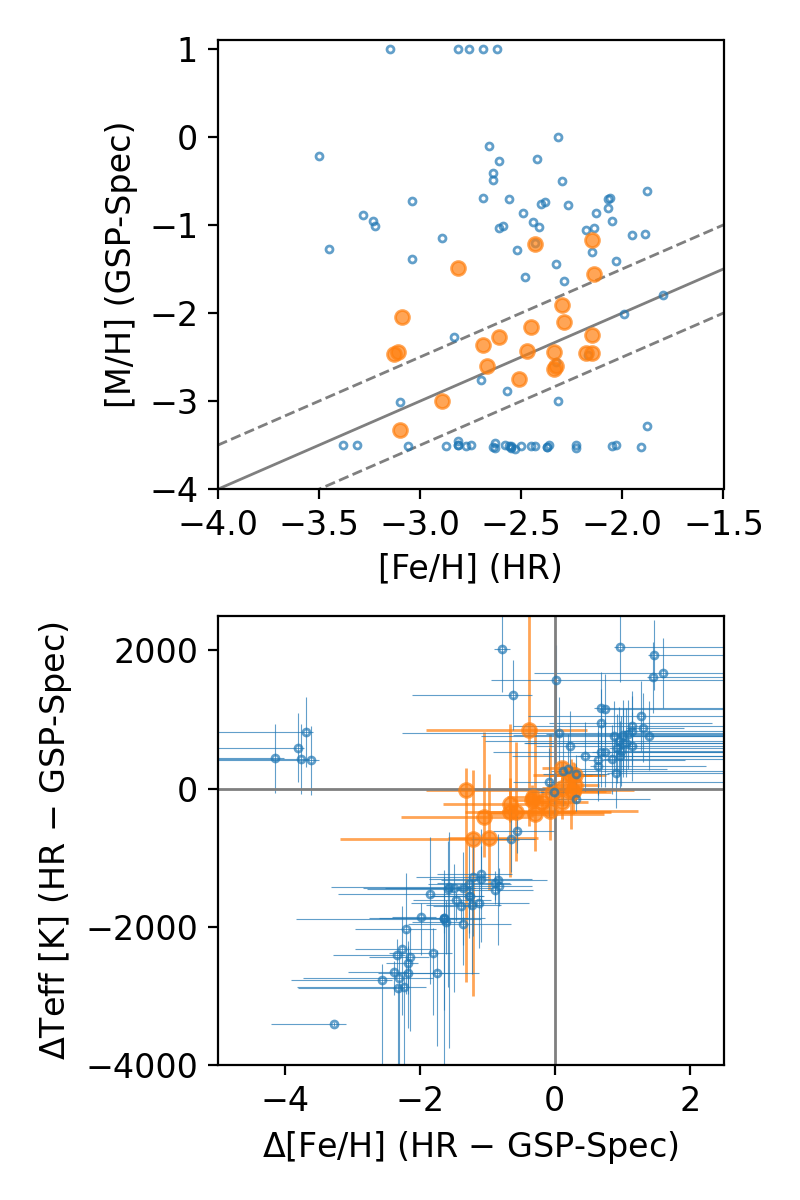}
  \end{subfigure}
  \caption{(Top:) Comparison of metallicities between the value from optical high-resolution spectroscopy and GSP-Spec metallicity for VMP stars from \citet{Li2022a}. (Bottom:) The relation between $T_{\rm eff}$ and [{Fe}/{H}] discrepancies. The orange points show stars that satisfy the filters introduced by \citet{RecioBlanco2022a} for constructing the metallicity distribution function. \label{fig:gspspec_VMP}}
\end{figure}

Low-metallicity stars provide us with unique opportunities to study astrophysical processes \citep{Beers2005,Frebel2015}, such as the formation and supernovae explosion of first stars \citep[e.g.,][]{Ishigaki2018a}, nucleosynthesis of heavy elements by neutron-capture processes \citep[e.g.,][]{Sneden2008}, and the early assembly of the Milky Way \citep[e.g.,][]{Chiba2000a,Yuan2020a,Sestito2021a}.
Therefore, extensive efforts have been devoted to searching for such stars from photometric observations \citep{Schlaufman2014,Starkenburg2017a,DaCosta2019a} and spectroscopy \citep{Beers1985,Beers1992,Christlieb2008,Caffau2013a,Aoki2013,Roederer2014,Aguado2016,Li2018b,Matijevic2017a}. 
Since very metal-poor stars (VMP; $[\mathrm{Fe/H}]<-2$) are relatively rare among field stars, it is necessary both to observe a large number of stars and to develop an efficient method to select VMP stars out of them.

A new opportunity arrives thanks to the Gaia mission \citep{GaiaCollaboration2016a}.
For the first time, the recent Gaia Data Release \citep{GaiaCollaboration2022a} has provided stellar parameters and chemical abundances estimated from the spectra taken with the Radial Velocity Spectrometer \citep[RVS;][]{Cropper2018a} instrument.
The spectra cover the Ca triplet region ($846-870\,\mathrm{nm}$) with a resolution of $R\sim 11,500$, and are analyzed through the General Stellar Parametriser-spectroscopy, GSP-Spec module \citep{RecioBlanco2022a}. 
The current GSP-Spec catalog contains 5.6 million stars analyzed and is already one of the largest catalogs of stellar parameters derived from spectra; moreover, the number is expected to increase in future releases.
By filtering out stars with suspicious solutions, \citet{GaiaCollaboration2022b} demonstrate that the estimated astrophysical quantities have excellent quality.

Although the wavelength coverage of the RVS spectra is known to be suitable for searching for VMP stars \citep[e.g.,][]{Starkenburg2010a,Matijevic2017a}, there is a limitation in the current GSP-Spec parameters for this purpose.
The main difficulty is the small number of lines detected in the spectra of VMP stars \citep{Kordopatis2011a,RecioBlanco2016a}.
In the fully spectroscopic approach, which is adopted in the current version of the GSP-Spec module, one needs to simultaneously determine at least three parameters, namely effective temperature ($T_{\rm eff}$), surface gravity ($\log g$), and metallicity ([M/H])\footnote{The GSP-Spec module additionally varies $[\mathrm{\alpha/Fe}]$ in this step.}. 
Although this is possible for solar metallicity stars, 
\added{it gets more challenging for VMP stars, whose spectra lack spectral signatures needed for parameter determination} \citep{RecioBlanco2022a}.
Therefore, a number of filtering needs to be applied when studying metal-poor stars in the GSP-Spec catalog, and the precision in [M/H] is not as good as for solar metallicity stars even after the filtering. 
\added{The filtering process removes stars with potential parameter bias due to large broadening velocity and large radial velocity uncertainty, stars whose parameter solution is outside of the training grid, hot stars with $T_{\rm eff}>6000\,\mathrm{K}$, those classified as O or B type stars by the Extended Stellar Parameteriser of Hot Stars (ESP-HS), cool giants (K and M-type giants), and stars with high gravity ($\log g>4.9$).}
\added{\citet{RecioBlanco2022a} used this set of filters to study the metallicity distribution function, which we call MP-filter throughout this paper.}

\added{The difficulty is illustrated in Figure~\ref{fig:gspspec_VMP}, where we compare GSP-Spec metallicity with \citet{Li2022a}, who studied $\sim 400$ VMP stars with high-resolution spectroscopy. 
These VMP stars are originally selected from the low-resolution survey ($R\sim 1800$) of Large Sky Area Multi-Object Fibre Spectroscopic Telescope \citep[LAMOST;][]{Cui2012a,Zhao2012} using the method described by \citet{Li2018b} for the high-resolution follow-up observations with the Subaru telescope as described by \citet{Aoki2022a}.
Since the metallicity estimates by \citet{Li2022a} are based on high-resolution spectroscopy and homogeneous analysis, we can consider them reliable and precise. 
There are 109 stars in this sample that have been analyzed by the GSP-Spec module based on the RVS spectra with $S/N\sim 20-110$.}

\added{
Figure~\ref{fig:gspspec_VMP} shows that the GSP-spec module struggles to provide accurate metallicities at very low metallicity.
Not only the GSP-Spec metallicities have large uncertainties, which likely reflect strong parameter degeneracy, there is also a large population of stars for which GSP-Spec systematically overestimates metallicity and underestimates its uncertainty (see the third quadrant of the bottom panel).
Although the MP filter removes most of such cases, only 23 out of 109 stars (21\%) survive the filtering.
}

\begin{table}
\caption{The number of VMP stars removed by each condition\label{tab:MPfilterVMP}}
\centering
\begin{tabular}{lr}
\hline\hline 
$T_{\rm eff,gspspec}>6000$  & 43 \\
$KMgiantPar>0$ & 42 \\
\makecell[l]{$T_{\rm eff,gspspec}<3500$ or \\ ($T_{\rm eff,gspspec}<4150$ and $2.4<\log g<3.8$)} & 0 \\
$\log g_{\rm gspspec} >4.9$ & 27 \tablefootmark{a}\\
vboard[TGM]>1 & 4 \tablefootmark{b} \\
vrad[TGM]>1 & 20 \tablefootmark{c} \\
extrapol>2 & 0 \\
spectraltype\_esphs = 'O' or 'B' & 0\\
\hline
\end{tabular}
\tablefoot{This table is for 109 stars in common between \citet{Li2022a} and the GSP-Spec catalog. In total, 86 stars were removed due to these conditions introduced by \citet{RecioBlanco2022a}.
\tablefoottext{a}{All 27 stars also have $T_{\rm eff,gspspec}>6000$.}
\tablefoottext{b}{All 4 stars also have $KMgiantPar>0$.}
\tablefoottext{c}{Among the 20 stars, 14 stars also satisfy $T_{\rm eff,gspspec}>6000$ and 5 stars have $KMgiantPar>0$.}
}  
\end{table}

\added{The bottom panel of Figure~\ref{fig:gspspec_VMP}, however, shows a possibility of improving the metallicity agreement for stars removed by the MP filter.
There is a clear sequence between metallicity and temperature discrepancies among many of the VMP stars, indicating that most of the poor solutions are due to the degeneracy between [$\mathrm{Fe/H}$] and $T_{\rm eff}$.
If we are able to shift $T_{\rm eff}$ closer to the high-resolution $T_{\rm eff}$, we would expect the metallicity also to be shifted closer to the high-resolution value.}
\added{In this work, we incorporate photometric and astrometric information of stars into the analysis, in order to provide more precise and accurate metallicity for VMP stars.
We also aim to keep more VMP stars even after quality cuts by dropping or relaxing some of the criteria used in the MP filter, in particular, $T_{\rm gspspec}<=6000$ (and $\log<=4.9$) and $KMgiantPar$, which are responsible for removing many VMP stars (see Table~\ref{tab:MPfilterVMP}).
These two criteria were introduced to remove very hot OBA-type stars, which can be confused with VMP stars due to the current GSP-Spec grid border at $T_{\rm eff}=8000\,\mathrm{K}$, and cool giants, which can look metal-poor due to pseudo-continuum caused by molecular absorption. 
While it was difficult to separate FGK-type VMP stars from these contaminants using the current $T_{\rm gspspec}$ due to the parameter degeneracy, we expect that the separation is clear when using photometric and astrometric information.}
As spectra are publicly available only for 18\% of stars analyzed by the GSP-Spec module, we are not able to reanalyze spectra directly.
Instead, we assume that we can infer the observed spectra using the published GSP-Spec stellar parameters and remeasure metallicities from them. 
We describe this method in detail in Section~\ref{sec:method}.
\added{Using public RVS spectra and high-resolution surveys, we then validate the assumption, investigate the range where the assumption holds, and provide uncertainties to our new metallicities in Section~\ref{sec:validation}.
In Section~\ref{sec:discussion}, we present properties of VMP and extremely metal-poor (EMP; $[\mathrm{Fe/H}]<-3$) stars selected based on our new metallicity, and discuss possible caveats and future prospects.
} 
We present a summary in Section~\ref{sec:summary}.

\section{Method}\label{sec:method}
\begin{figure*}
\centering
  \begin{subfigure}[t]{0.95\textwidth}
  \centering
  \includegraphics[width=\linewidth]{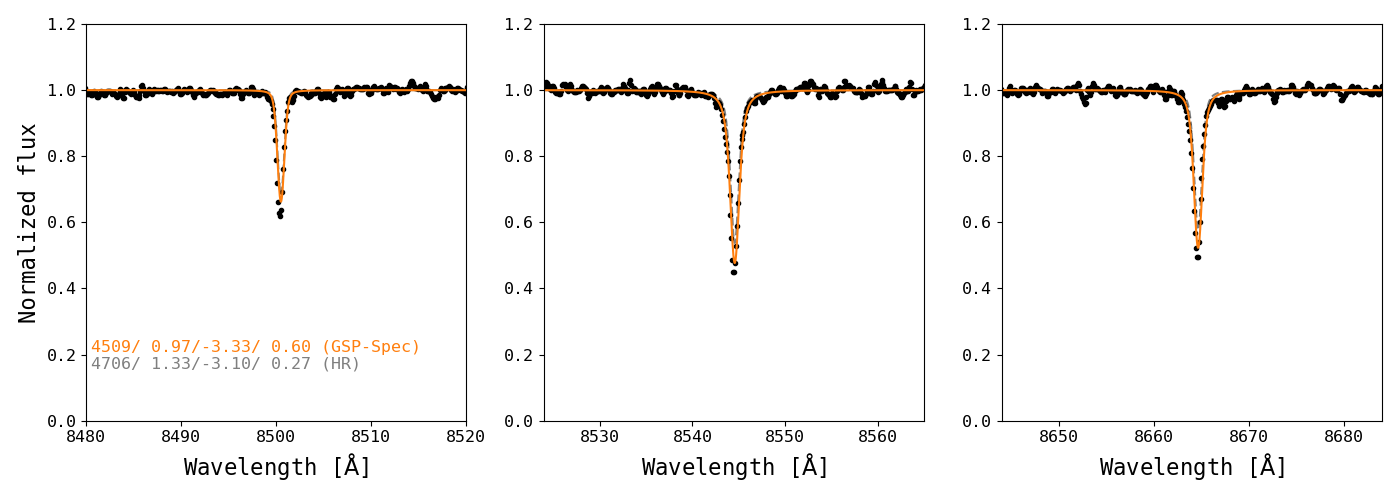}
  \end{subfigure}
  \begin{subfigure}[b]{0.95\textwidth}
  \centering
  \includegraphics[width=\linewidth]{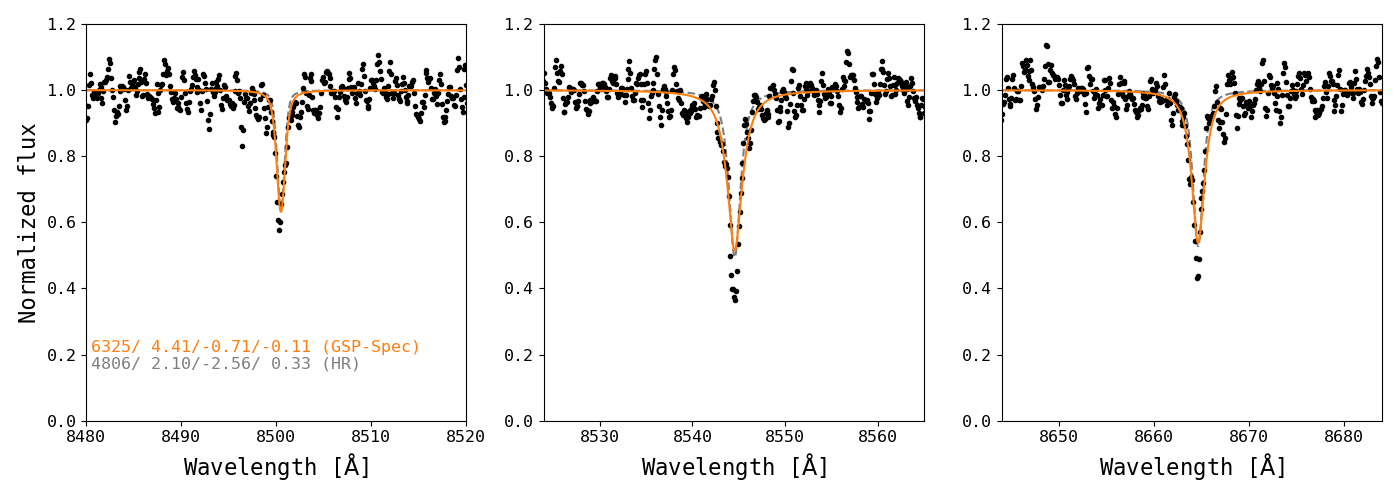}
  \end{subfigure} 
  \caption{Comparison of observed RVS spectra (shown in black dots) with synthetic spectra (orange solid and gray dashed lines). The orange and gray spectra are reproduced with GSP-Spec and \citet{Li2022a} parameters, respectively. The latter study is based on high-resolution spectroscopy (HR). The top and bottom panels are for a star that satisfies the MP filters (Gaia DR3 927340858625174656) and for one that does not (Gaia DR3 2682719929806900096). The parameters assumed in the syntheses are shown in panels in the order of $T_{\rm eff}$, $\log g$, [{Fe/H}], and [$\alpha$/{Fe}] ([{Ca}/{Fe}] in case of HR). Our approach is to infer a new metallicity based on the orange spectrum since not every star in the GSP-Spec catalog has a publicly available observed spectrum. \label{fig:spec}}
\end{figure*}

\subsection{Photometric $T_{\rm eff}$ and $\log g$}

We here provide independent $T_{\rm eff}$ and $\log g$ for stars in the current GSP-Spec catalog using photometry and astrometry.
Since these stellar parameters do not rely on RVS spectra, we expect them to help us break the degeneracy among parameters.

We use Gaia and the Two Micron All Sky Survey \citep[2MASS][]{Skrutskie2006a} as the source of photometric information.
We first crossmatch stars analyzed by the GSP-Spec module with Gaia main source catalog and 2MASS point source catalog \citep{Cutri2003a} using the 2MASS best neighbor catalog in Gaia DR3 (\texttt{gaiadr3.tmass\_psc\_xsc\_best\_neighbour}). 
We primarily use three-dimensional extinction maps by \citet{Green2019a} and \citet{Marshall2006a} to deredden the photometries.
For stars that are not covered by either of the three-dimensional maps, we scale the two-dimensional map by \citet{Schlegel1998a} assuming a $\sech ^2$ profile for dust distribution, where the scale height was assumed to be $125\,\mathrm{pc}$ \citep{Marshall2006a}.
The extinction laws are taken from the values presented in \citet{Casagrande2021a} for \citet{Cardelli1989a} and \citet{ODonnell1994a} extinction laws.

We estimate the photometric $T_{\rm eff}$ using $G-K_s$ color when 2MASS $K_s$-band magnitude has a quality `A' or `B'; otherwise, we use Gaia $BP-RP$ color.
We adopt color-temperature relations by \citet{Mucciarelli2021a}, \added{around which there are about $50-60\,\mathrm{K}$ dispersions.}
We adopt the nearest edge value of the range for stars with metallicity values outside the calibration range.
We discard stars falling outside the color range in which the relations are calibrated.

We then compute $\log g$ of stars from their luminosities ($L$).
Combined with the photogeometric distance by \citet{BailerJones2021a}, we obtain $L$ from $K_s$-band if it was measured with the aforementioned quality or $G$-band otherwise, using the bolometric correction of \citet{Casagrande2014,Casagrande2018a}\footnote{We use an updated version available at \url{https://github.com/casaluca/bolometric-corrections}. We also use our reimplementation of their python script for more efficient computation.}.
We then derive $\log g$ of stars from $g\propto M/R^2\propto MT_{\rm eff}^4/L$, where the stellar mass $M$ was assumed to be $0.7\,\mathrm{M_\odot}$ for stars with $\log g>3.5$ and $0.8\,\mathrm{M_\odot}$ for $\log g<3.5$. 
Although the assumption of mass can be inaccurate for some stars, it does not affect our conclusion since we are interested in metal-poor stars, which are known to be old. 
In addition, the effect of $M$ is minimized when converting the gravity to the log scale for constructing model atmospheres. 

\added{
We estimate the typical uncertainty of $\log g$ as follows.
If a star has $20\%$ distance fractional uncertainty, it contributes to $0.17\,\mathrm{dex}$ $\log g$ uncertainty.
However, since the GSP-Spec stars tend to have precise distance, with the median fractional distance uncertainty of $1.9\%$, the typical contribution from the distance is only $0.016\,\mathrm{dex}$.
The $T_{\rm eff}$ does not contribute to the $\log g$ uncertainty either; $60\,\mathrm{K}$ uncertainty in $T_{\rm eff}$ is translated to $0.017$ uncertainty in $\log g$.
The dominant source of $\log g$ uncertainty is the assumed mass.
For low metallicity stars, stars usually have mass within $0.1\,\mathrm{M_{\odot}}$ form the assumed mass, which would be translated to $0.05-0.07\,\mathrm{dex}$ $\log g$ uncertainty.
The effect of mass can be larger for high metallicity stars, which can be young.
For example, if a star has $1.6\,\mathrm{M_{\odot}}$ instead of $0.8\,\mathrm{M_{\odot}}$ assumed in the analysis, we would underestimate $\log g$ by $0.3\,\mathrm{dex}$.
}

\subsection{Remeasuring metallicity\label{sec:method_feh}}

The RVS spectra are usually dominated by the Ca triplet feature. 
We, therefore, assume that the GSP-Spec solution provides a reasonable fit to the Ca triplet even when the solution suffers from degeneracy.
This assumption is justified by the distribution of $\log \chi^2_{\rm gspspec}$, which quantifies the residual between the best-fit spectrum and the observed spectrum.
The median $\log \chi^2_{\rm gspspec}$ are $-3.06$ and $-2.93$ for the stars in common between the MP sample and \citet{Li2022a} and for the remaining stars from \citet{Li2022a}, indicating that the goodness of the fits does not depend much on whether the star satisfies the MP-filters.\footnote{We remove stars with $T_{\rm eff,gspspec}=4250,\mathrm{K}$ and $\log g_{\rm gspspec}=1.5$ in this comparison since the MP filters have a cut in $KMgiantPar$, which depends on $\log \chi^2_{\rm gspspec}$ for stars that have these stellar parameters in the published catalog \citep{RecioBlanco2022a}. Such a cut can introduce a bias in the $\log \chi^2_{\rm gspspec}$.}
We can also visually confirm the assumption in Figure~\ref{fig:spec}, which compares published RVS spectra with synthetic spectra for two stars selected from \citet{Li2022a}.
\added{We calculated the two synthetic spectra for the published GSP-Spec parameters and the high-resolution parameters and abundances in a consistent way as the GSP-Spec module; we use Turbospectrum v19.1 \citep{Plez2012a} with MARCS model atmospheres \citep{Gustafsson2008a} and the same line data for the Ca triplet presented in \citet{Contursi2021a}.
For the microturbulence, we assume the relation in \citet{Buder2021a} for the GALAH survey since that used by the GSP-Spec module is not publicly available.
}
The good match among the three spectra, regardless of how well the GSP-Spec parameters agree with high-resolution values, is encouraging. 
\added{We further validate the assumption in Section~\ref{sec:ew_validation}.}

Based on this assumption, we first compute Ca triplet equivalent widths from the published values of stellar parameters and then rederive Ca abundance of stars using the $T_{\rm eff}$ and $\log g$ from the previous section \added{using the software, model atmospheres, line data and microturbulence described above}. 
For computational reasons, we adopt interpolation in a pre-constructed grid for the analysis.
The grid provides equivalent widths for the three Ca lines as a function of $T_{\rm eff}$, $\log g$, and [{Ca}/{H}].
We only use models with standard chemical compositions, which have the following relation between [$\alpha$/{M}] and [{M}/{H}]:
\begin{align}
[\mathrm{\alpha/Fe}]=
\begin{cases}
0.4 & ([\mathrm{Fe/H}] \leq -1.0) \\ 
-0.4[\mathrm{Fe/H}] & (-1.0< [\mathrm{Fe/H}]< 0.0)\\
0.0 & (0\leq [\mathrm{Fe/H}]) 
\end{cases}.\label{eq:alphafe}
\end{align}
The [{Ca}/{H}] in our grid is equal to [$\alpha$/{H}].

As the first step, we estimate the equivalent widths of the Ca triplet using GSP-Spec stellar parameters, $T_{\rm eff}$, $\log g$, [{M}/{H}], and [$\alpha$/M].
We convert the reported [{M}/{H}] and [$\alpha$/{M}] to [$\alpha$/{H}] and take this value as the input [{Ca}/{H}].
Since we aim to mitigate the difficulty in the global fitting of spectra, we use the global stellar parameters rather than using abundances of individual elements such as [{Ca}/{Fe}].
In this process, GSP-Spec only varies the four parameters, not individual abundances.
When a star does not have a reported [$\alpha$/M], we assume the Equation~\ref{eq:alphafe} to convert [{M}/{H}] to [{Ca}/{H}]. 
For stars with $T_{\rm eff}>4000$ and $\log g >5.0$, we adopt $\log g=5.0$ as these are not represented in our model grid.

Using the adopted $T_{\rm eff}$ and $\log g$ from the previous section, we measure Ca abundance from each of the three lines from the estimated equivalent widths.
We take the median of the three estimates for the adopted value of [{Ca}/{H}].
We finally convert the Ca abundance to metallicity assuming Equation~\ref{eq:alphafe}.
Since both the photometric temperature and bolometric correction require a pre-determined metallicity, we start with the GSP-Spec metallicity and iterate the whole process three times to ensure convergence.

\section{Validation}\label{sec:validation}

\added{In this section, we validate our assumption on the recovery of the Ca triplet equivalent widths and compare our new metallicities with previous studies.
Through these validations, we will introduce new quality cuts and estimate the typical uncertainty in our new metallicity.
}

\subsection{The recovery of equivalent widths \label{sec:ew_validation}}

\added{We here investigate how well we can recover the equivalent widths of the Ca triplet based on the assumption in Section~\ref{sec:method_feh} by comparing the recovered values with the values directly measured from spectra.
We use public RVS spectra, which are available for $\sim 18\%$ of the stars in the GSP-Spec catalog.
Since we are mostly interested in low-metallicity stars, we downloaded all the public RVS spectra for 1934 stars which have $[\mathrm{Fe/H}]<-2$ in our new metallicity.
We then fit a Voigt profile to each of the three Ca triplet lines to measure equivalent widths.
}

\begin{figure*}
\includegraphics[width=\textwidth]{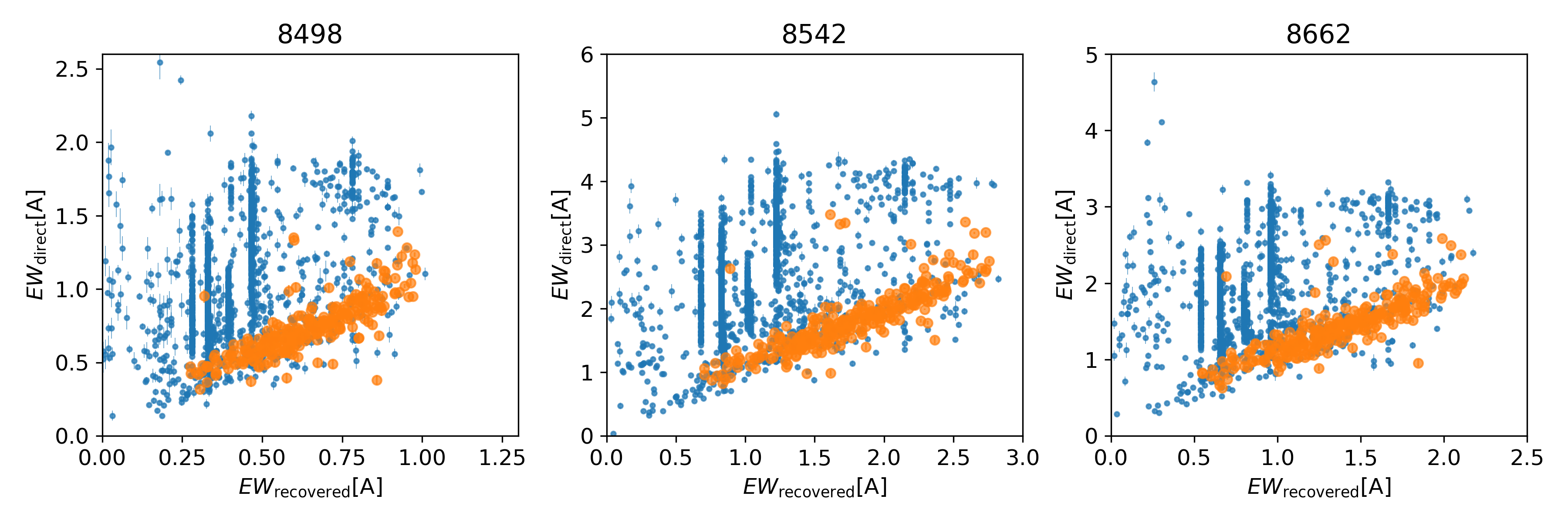}
\caption{Comparisons of the equivalent widths recovered based on the assumption in Section~\ref{sec:method_feh} and those directly measured from spectra. Symbols follow Figure~\ref{fig:gspspec_VMP}. \label{fig:ew_compare}}
\end{figure*}

\begin{table*}
  \caption{Eequivalent widths comparisons \label{tab:ew_compare}}
  \centering
  \begin{tabular}{lrrrrrl}
  \hline\hline
   & \multicolumn{3}{c}{$\langle \Delta \log EW\rangle$} &  Adopted & Adopted & \\ \cline{2-4}
  Sample \tablefootmark{a} & median & 16th percentile & 84th percentile & correction & uncertainty & comment\\\hline
    MP sample  &  0.02 &  -0.01 & 0.08 & 0.00 & 0.05 \\
    A          &  0.05 &  -0.02 & 0.10 & 0.00 & 0.05 & added in RMP\\
    B          &  0.05 &  -0.02 & 0.10 & 0.00 & 0.05 & added in RMP\\ 
    E1         &  0.10 &  -0.01 & 0.26 & -0.05 & 0.13  & added in eRMP1\\ 
    E2         &  0.20 &  0.12  & 0.30 & -0.15 &  0.14 & added in eRMP2\\
  \hline
  \end{tabular}
  \tablefoot{
  \tablefoottext{a}{See Figures~\ref{fig:ew_compare} and \ref{fig:dlogew_KM} and texts for the definition of the samples.}
  }
\end{table*}

\added{Figure~\ref{fig:ew_compare} compares the recovered and directly-measured equivalent widths.
We here define $\langle \Delta \log EW\rangle$ as $\log (EW_{\rm recovered}/EW_{\rm direct})$ averaged over the three lines.
The good agreement between the two equivalent widths is clear among the MP sample; 16, 50, and 84 percentiles in $\langle \Delta \log EW \rangle$ are $-0.01$, $0.02$, and $0.08$ (Table~\ref{tab:ew_compare}).
}

\added{
However, the dispersion is significantly larger than the fitting uncertainty in $EW_{\rm direct}$, which is typically $\sigma(\log EW)<0.01\,\mathrm{dex}$.
Therefore, the dominant source of the dispersion should be due to what is not accounted in the uncertainty estimates in $EW_{\rm direct}$.
A possible source of the dispersion is the error due to our assumption that we can recover the equivalent widths from the published GSP-Spec parameters.
In particular, we do not account for the uncertainties in the GSP-Spec parameters. 
However, considering the GSP-Spec parameters' uncertainties turns out to be not straightforward because we do not have access to the full Monte Carlo sampling of GSP-Spec parameters nor the covariance among the parameters, which is crucial when estimated parameters suffer from significant degeneracy.
From error propagation, we have the following relation between the uncertainty in the recovered equivalent widths and stellar parameters' uncertainties:}
\begin{equation}
\displaystyle \sigma^2(EW_{\rm recovered})=\sqrt{\sum_i (\frac{\partial EW}{\partial \xi_i})^2\sigma^2(\xi_i)+\sum_{i\neq j}(\frac{\partial EW}{\partial \xi_i}\frac{\partial EW}{\partial \xi_j}\sigma_{\xi_i,\xi_j})},\label{eq:sigma_ew}
\end{equation}
\added{where $\xi_i$ is a stellar parameter, and $\sigma_{\xi_i,\xi_j}$ is the covariance between two stellar parameters.
Since we do not have access to $\sigma_{\xi_i,\xi_j}$, we have to assume them to be $0$.
Under this assumption, the median $\sigma(EW_{\rm recovered})$ is $\sim 0.15-0.16\,\mathrm{dex}$ for the MP sample, which is significantly larger than the dispersion in Figure~\ref{fig:ew_compare} ($0.03-0.06\,\mathrm{dex}$). 
This overestimation would likely be due to the significant correlation between parameters; for example, while $\frac{\partial EW}{\partial T_{\rm eff,gspspec}}$ and $\frac{\partial EW}{\partial [\mathrm{Fe/H}]_{\rm gspspec}}$ being negative and positive, respectively, $\rho_{T_{\rm eff,gspspec},[\mathrm{Fe/H}]_{\rm gspspec}}=\sigma_{T_{\rm eff,gspspec},[\mathrm{Fe/H}]_{\rm gspspec}}/\sigma_{T_{\rm eff,gspspec}}\sigma_{[\mathrm{Fe/H}]_{\rm gspspec}}$ seems positive and significantly larger than 0 (see Figure~\ref{fig:gspspec_VMP}).
Thus, the term including $\sigma_{T_{\rm eff,gspspec},[\mathrm{Fe/H}]_{\rm gspspec}}$ reduces the estimated uncertainty.
Since we are not able to take this term into account, we should naturally expect to overestimate $\sigma(EW_{\rm recovered})$.
}

\added{We, therefore, conclude that it is not possible to estimate $\sigma(EW_{\rm recovered})$ from the currently available information. 
Thus we adopt an empirical approach to estimate the error caused by our assumption in Section~\ref{sec:method_feh}.
For the MP sample, we estimate $\sigma(\log EW)$ to be 0.05 from the dispersion in the comparison presented in Figure~\ref{fig:ew_compare}.
}

\begin{figure}
\centering
\includegraphics[width=\linewidth]{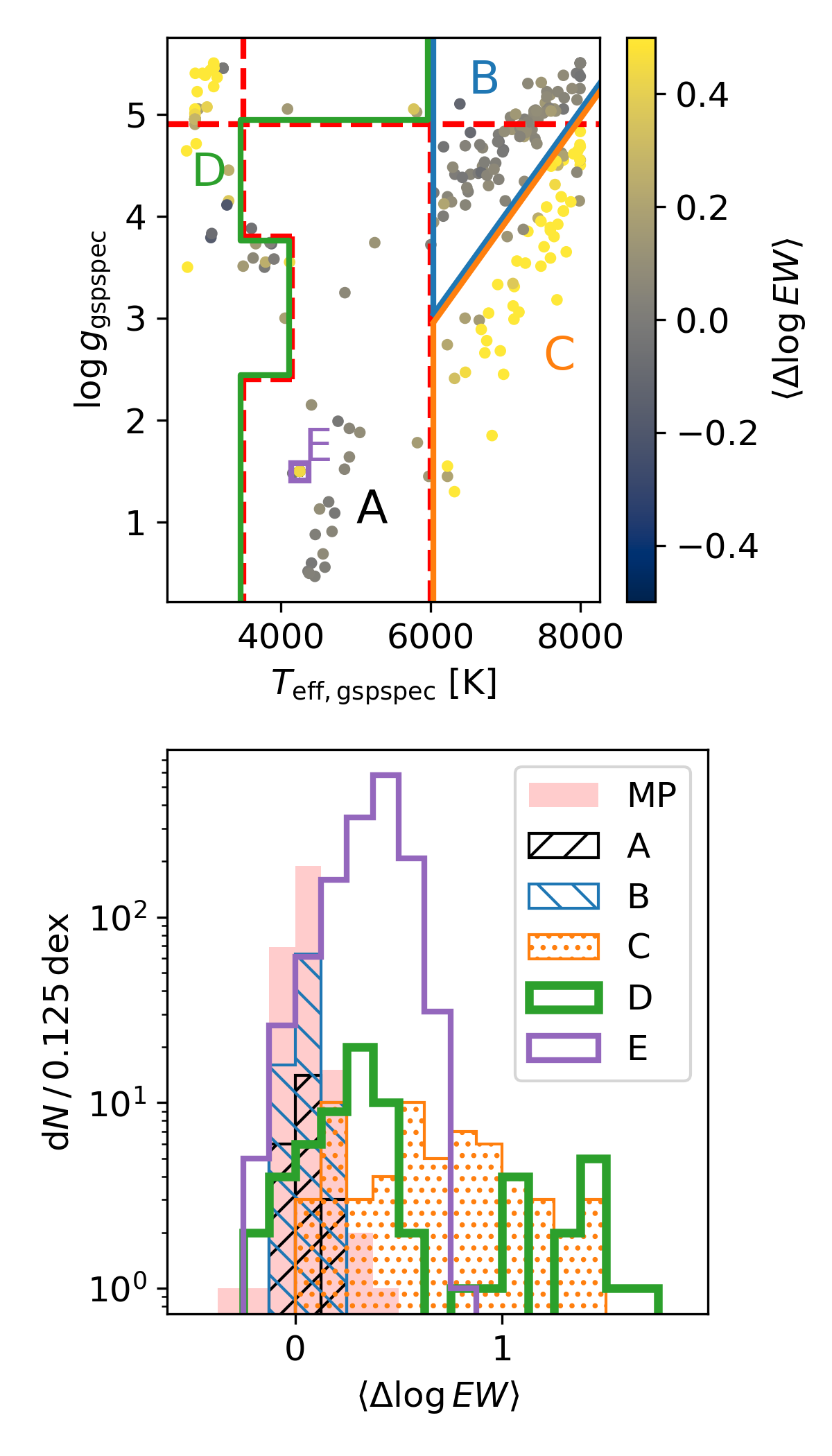}
\caption{The $\langle \Delta \log EW\rangle$ values across the Kiel diagram for stars that are filtered out by the MP filters. We define five regions A-E in the Kiel digram and investigate stars in each region. The boundary of region A is defined by the cuts made in MP filters \citep[][see Table~\ref{tab:MPfilterVMP}]{RecioBlanco2022a}. \label{fig:dlogew_kiel}}
\end{figure}
\begin{figure}
\centering
\includegraphics[width=\linewidth]{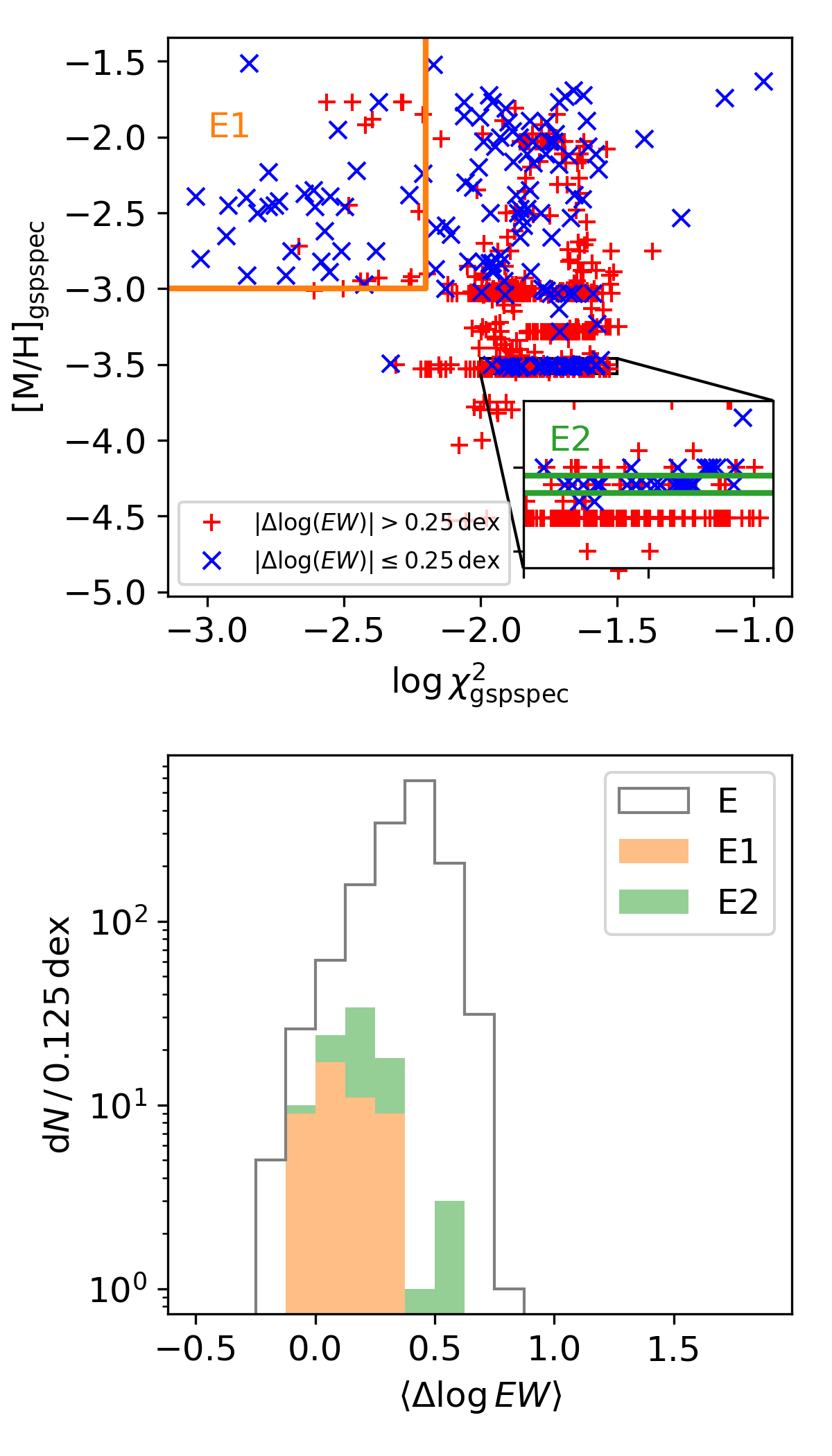}
\caption{The $\langle \Delta \log EW\rangle$ for stars with $KMgiantPar>0$. Top panel shows distribution of stars with a good agreement between recovered and directly measured equivalent widths (red) and those with a poor agreement (blue). We here define two regions E1 and E2 in $[\mathrm{Fe/H}]_{\rm gspspec}-\log\chi^2_{\rm gspspec}$ space. \label{fig:dlogew_KM}}
\end{figure}

\added{
Figure~\ref{fig:ew_compare} also shows that the Ca triplet equivalent widths are recovered for some of the stars that do not satisfy the MP filters.
Motivated by this, we here try to relax some of the quality cuts adopted by \citet{RecioBlanco2022a}, especially those effective in filtering out true VMP stars (Table~\ref{tab:MPfilterVMP}).
Figure~\ref{fig:dlogew_kiel} presents $\langle\Delta \log EW\rangle$ across the GSP-Spec Kiel diagram for stars that do not survive the MP filters.
We divide the stars into five boxes A-E depending on their positions in the diagram.}

\added{
The figure clearly shows that our method successfully recovers the equivalent widths for stars in regions A and B even though they do not satisfy the MP filters.
Stars in region A of Figure~\ref{fig:dlogew_kiel} survive all the cuts on stellar parameters in the MP filter but are removed by the other cuts, more specifically, vbroad[TGM]>1 and vrad[TGM]>1.
The good recovery for the stars in the region B opens a possibility to retain more warm metal-poor stars, which are originally removed by requiring $T_{\rm eff, gspspec}<6000\,\mathrm{K}$ in the MP filters (see Table~\ref{tab:MPfilterVMP}).
For stars in regions A and B, the (16, 50, and 84) percentiles of $\langle\Delta\log EW\rangle$ are ($-0.02$, $0.05$, $0.10$) and ($-0.02$, $0.05$, $0.10$), respectively (Table~\ref{tab:ew_compare}).
These values are similar to what we found for the MP sample.
On the other hand, our method fails to recover the Ca triplet equivalent widths for the majority of the stars in regions C and D.
}

\added{
We discuss region E separately here. 
The region E corresponds to $T_{\rm eff,gspspec}=4250\,\mathrm{K}$ and $\log g_{\rm gspspec}=1.5$, and all the stars in this region have $KMgiantPar>0$.
Any improvement for this subsample is critical in retaining a larger number of metal-poor stars (see Table~\ref{tab:MPfilterVMP}).
While the equivalent widths recovery for them seems challenging in general, an additional space, defined by $[\mathrm{Fe/H}]_{\rm gspspec}$ -- $\log\chi^2_{\rm gspspec}$, opens a possibility to rescue some stars in this category by making a further cut.
Figure~\ref{fig:dlogew_KM} shows that the fraction of stars with a good agreement between direct measurements and recovered values changes as a function of $\log \chi^2_{\rm gspspec}$ and $[\mathrm{Fe/H}]_{\rm gspspec}$.
In particular, the agreement seems reasonably good for two regions, the one (E1) defined as $[\mathrm{Fe/H}]_{\rm gspspec}>-3$ and $\log \chi^2_{\rm gspspec}<-2.0$ and the other (E2) defined as $[\mathrm{Fe/H}]_{\rm gspspec}=3.51$.
For stars in regions E1 and E2, the (16, 50, and 84) percentiles of $\langle\Delta\log EW\rangle$ are ($-0.01$, $0.10$, $0.26$) and ($0.12$, $0.20$, $0.30$), respectively (Table~\ref{tab:ew_compare}). 
}

\subsection{Relaxing filters and uncertainties}\label{sec:uncertainty}
\begin{figure*}
  \centering
  \includegraphics[width=\textwidth]{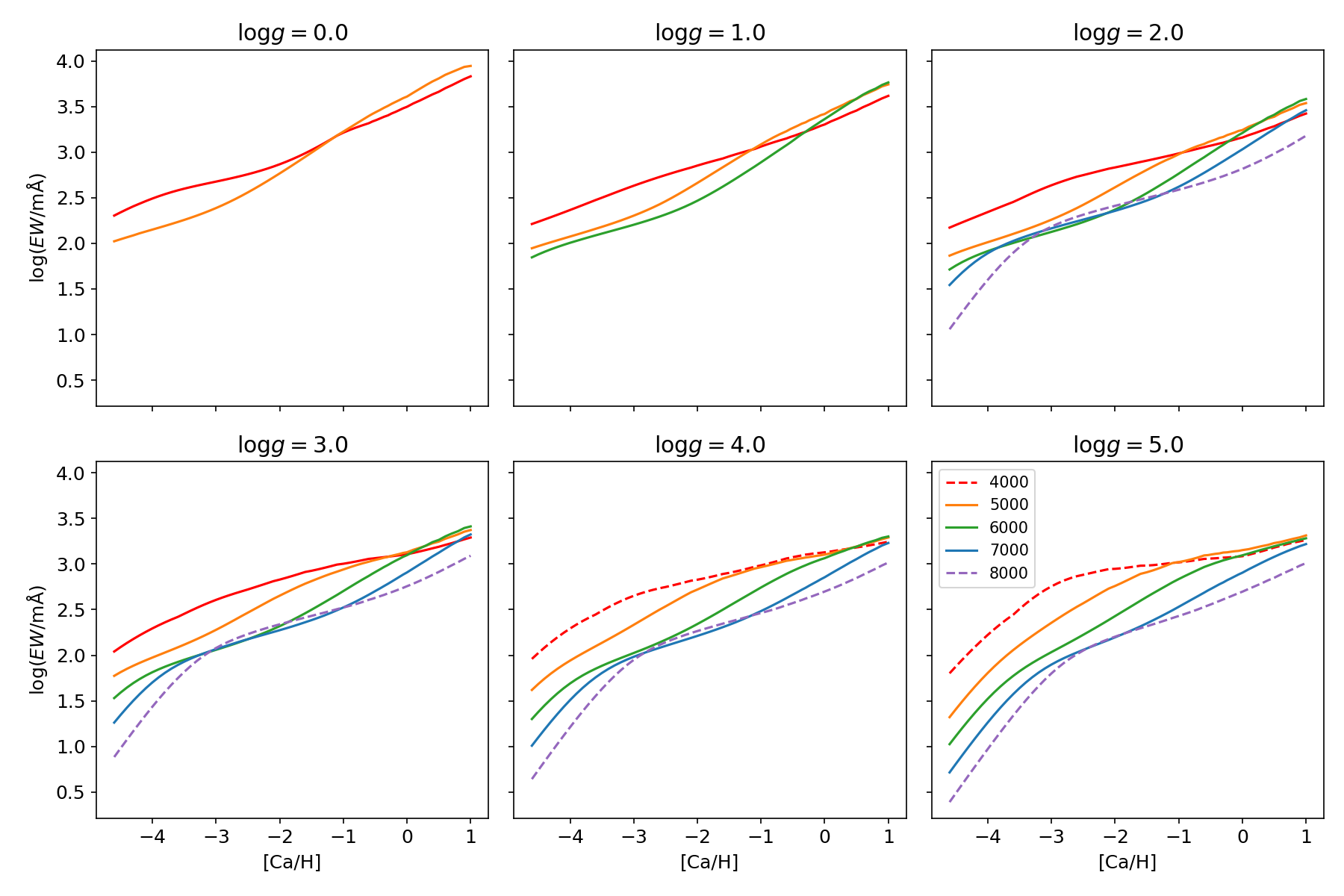}
  \caption{The curve of growth for the Ca line at $8498\,\mathrm{\AA}$. Colors show different $T_{\rm eff}$. Dashed lines are used for $T_{\rm eff, phot}>7000\,\mathrm{K}$, and $T_{\rm eff, phot}<5000\,\mathrm{K}$ and $\log g_{\rm phot}>3.0$ as they are not included in the final sample since the Ca lines can be contaminated by Hydrogen lines in too hot stars and their strengths are not sensitive to the Ca abundance in cool dwarfs (see the two right panels in the bottom row). Similar figures for the other two lines are available in Appendix~\ref{appendix:cog}.}\label{fig:sensitivity}
\end{figure*}

\added{Based on the validation presented in the previous subsection, we here define four types of quality cuts and present a typical uncertainty in our newly derived metallicity for each sample.
Before relaxing the quality cuts, we here introduce additional quality cuts to remove stars for which our method is expected to perform poorly.
Since we rely on photometry and astrometry, the precision of stellar parameters degrades for stars with high extinction or poor distance estimation.
Our method also suffers in hot stars and cool dwarfs since our assumption that the Ca triplet dominates the RVS spectra does not hold for such stars. 
While Hydrogen lines start contaminating the Ca triplet in hot stars, the Ca triplet strengths are no longer sensitive to the metallicity in cool dwarfs (see Figure~\ref{fig:sensitivity}).
To mitigate all these effects, we introduce additional filters on extinction ($E(B-V)<0.72$), distance precision ($2d_{\rm photogeo,med}/(d_{\rm photogeo,hi}-d_{\rm photgeo,lo})>3$), and on our photometric stellar parameters ($T_{\rm eff}<7000\,\mathrm{K}$, and $T_{\rm eff}>5000\,\mathrm{K}$ or $\log g<3.0$).
}

\added{The first sample combines these additional quality cuts and the MP filters, which we call MP+ sample.
Some of the conditions are then dropped or relaxed to form RMP sample following the investigation of the previous section. 
Specifically, we drop the conditions on vbroad[TGM] and vrad[TGM], and relax the condition on stellar parameters to include stars in region B of Figure~\ref{fig:ew_compare}.
We further extend the sample by including stars in sample E1 of the previous section, which has $KMgiantPar=1$, $[\mathrm{M/H}]_{\rm gspspec}>-3$, and $\log\chi^2_{\rm gspspec}<-2.2$ (eRMP1 sample), and then those in E2, which have $KMgiantPar=1$ and $[\mathrm{Fe/H}]_{\rm gspspec}=-3.51$ (eRMP2 sample).
These sets of quality cuts and samples are summarized in Appendix~\ref{appendix:qc}.
}

\added{For the stars in MP+ and RMP samples, we adopt $\sigma(\log EW)=0.05\,\mathrm{dex}$ as the uncertainty in the recovered equivalent widths (Table~\ref{tab:ew_compare}). 
Note that we ignored the small offset between $EW_{\rm recovered}$ and $EW_{\rm direct}$ since the offset might just be due to the mismatch between observed and synthetic spectra at the cores of the Ca triplet.
Since we use the same approach when recovering equivalent widths and remeasuring metallicities, we consider that the effect of the offset would be canceled out.
We apply corrections to the recovered equivalent widths for stars newly added in the eRMP1 and eRMP2 samples, as the offsets found for these samples are different from what we found for the MP sample.
We shift the recovered equivalent widths by $-0.05$ and $-0.15$ dex in $\log(EW)$ for these stars, respectively, based on the difference in the median $\langle \Delta\log EW\rangle$ values between the MP and eRMP samples (Table~\ref{tab:ew_compare}).
Similarly, we adopt $0.13$ and $0.14$ dex uncertainties for them.
}

\added{
Taking into account the typical uncertainties in our stellar parameters, $T_{\rm eff}$ ($60\,\mathrm{K}$) and $\log g$ ($0.08\,\mathrm{dex}$), and those in the recovered equivalent widths, we estimate the uncertainty in remeasured metallicity for individual objects.
The typical metallicity uncertainties (the number of objects) are 0.32 (3,134,736), 0.30 (4,109,821), 0.30 (4,142,936), and 0.30 (4,143,310) dex for MP+, RMP, eRMP1, and eRMP2 samples.
In all cases, the contribution of $\sigma(\log EW)$ is by far the largest;
even for MP+ sample, $\sigma(\log EW)$ has $\sim 5$ and $\sim 10$ times larger contributions than $\sigma(T_{\rm eff})$ and $\sigma(\log g)$, respectively.
We note, however, that these uncertainties are estimated using stars with $[\mathrm{Fe/H}]<-2$ in our new metallicity estimates. 
As we will see in the next section, higher metallicity stars seem to have smaller uncertainties.
}

\begin{figure*}
  \includegraphics[width=\textwidth]{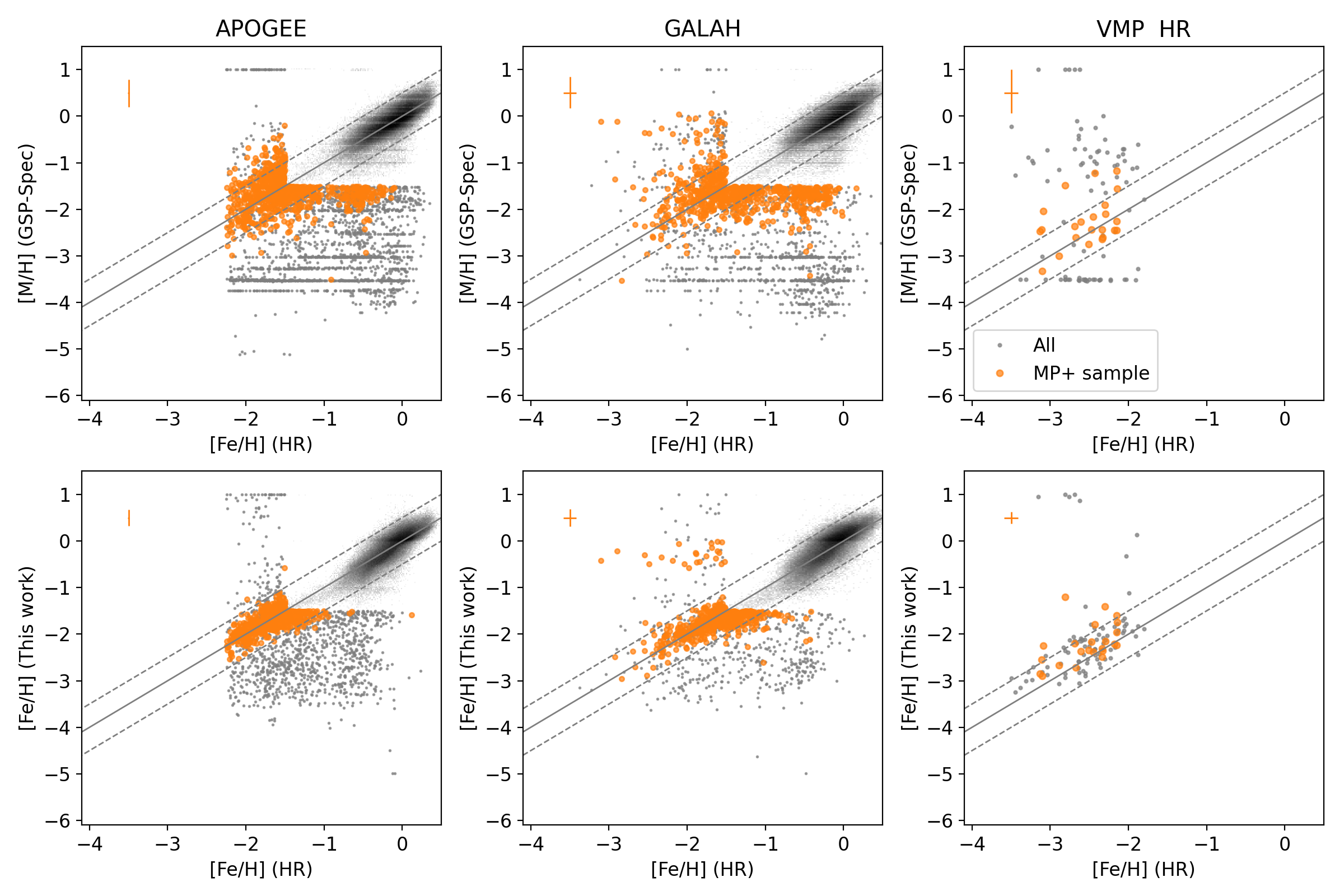}
  \caption{Comparisons between [{Fe}/{H}] from surveys and our new metallicity estimates. All stars with $[\mathrm{Fe/H}]<-1.5$ in either axis are shown as individual data points. In all panels, gray points show all the stars available for the comparison, while the orange stars are for the MP+ sample, which combines the filters from \citet{RecioBlanco2022a} and our additional quality cuts in photometry, astrometry, and our photometric stellar parameters (see text for details). The solid and dashed lines respectively show one-to-one relation and $\pm 0.5$ dex offset. Median uncertainties are presented in the top left. \label{fig:validation}}
\end{figure*}

\subsection{Metallicity comparison with literature}\label{sec:feh_comparison}

\begin{table*}
  \caption{Metallicity comparisons with high-resolution studies\label{tab:result}}
  \centering
  \begin{tabular}{lrrrrl}
  \hline\hline 
   & \multicolumn{3}{c}{Median $\Delta [\mathrm{Fe/H}] \pm \mathrm{Std. dev}$ relative to surveys} & Median & \\\cline{2-4}
  Selection                             & APOGEE         & GALAH          & \citet{Li2022a}& $\sigma([\mathrm{Fe/H}])$\tablefootmark{a}  & [Fe/H] source \\\hline
  MP+\tablefootmark{c}                                   & $ 0.01\pm0.17$ & $ 0.06\pm0.19$ &                &   0.06                            & GSP-Spec \\
                                        & $-0.01\pm0.16$ & $ 0.06\pm0.19$ &                &   0.32                            & This work\\
  ($N$)                                 & 141298         & 104050         &                &                                   &          \\
  MP+ ($[\mathrm{Fe/H}]_{\rm HR}<-1.5$) & $ 0.15\pm0.36$ & $ 0.14\pm0.38$ & $0.13\pm0.46$  &   0.33                            & GSP-Spec \\
                                        & $ 0.02\pm0.17$ & $-0.01\pm0.19$ & $0.21\pm0.29$  &   0.18                            & This work\\
  ($N$)                                 & 571            & 324            & 22             &                                   &          \\\hline
  RMP \tablefootmark{d}                                  & $ 0.01\pm0.20$ & $ 0.06\pm0.22$ &                &   0.08                            & GSP-Spec \\
                                        & $ 0.00\pm0.17$ & $ 0.07\pm0.20$ &                &   0.30                            & This work\\
  ($N$)                                 & 174753         & 142535         &                &                                   &          \\
  RMP ($[\mathrm{Fe/H}]_{\rm HR}<-1.5$) & $ 0.16\pm0.43$ & $ 0.17\pm0.51$ & $1.22\pm0.88$  &   0.35                            & GSP-Spec \\
                                        & $ 0.02\pm0.18$ & $-0.01\pm0.20$ & $0.22\pm0.25$  &   0.18                            & This work\\
  ($N$)                                 & 654            & 390            & 64             &                                   &          \\\hline
  eRMP1 \tablefootmark{e}                                & $ 0.01\pm0.20$ & $ 0.06\pm0.22$ &                &   0.08                            & GSP-Spec \\
                                        & $ 0.00\pm0.17$ & $ 0.07\pm0.20$ &                &   0.30                            & This work\\
  ($N$)                                 & 175635         & 142606         &                &                                   &          \\
  eRMP1($[\mathrm{Fe/H}]_{\rm HR}<-1.5$)& $ 0.15\pm0.44$ & $ 0.15\pm0.52$ & $1.21\pm0.88$  &   0.35                            & GSP-Spec \\
                                        & $ 0.01\pm0.19$ & $-0.00\pm0.21$ & $0.23\pm0.34$  &   0.18                            & This work\\
  ($N$)                                 & 678            & 413            & 67             &                                   &          \\\hline
  eRMP2  \tablefootmark{f}                               & $ 0.01\pm0.20$ & $ 0.06\pm0.22$ &                &   0.08                            & GSP-Spec \\
                                        & $ 0.00\pm0.17$ & $ 0.07\pm0.20$ &                &   0.30                            & This work\\
  ($N$)                                 & 175667         & 142611         &                &                                   &          \\
  eRMP2($[\mathrm{Fe/H}]_{\rm HR}<-1.5$)& $ 0.14\pm0.55$ & $ 0.15\pm0.55$ & $1.02\pm1.06$  &   0.36                            & GSP-Spec \\
                                        & $ 0.02\pm0.21$ & $-0.00\pm0.22$ & $0.24\pm0.34$  &   0.18                            & This work\\
  ($N$)                                 & 704            & 418            & 76             &                                   &          \\
  \hline
  \end{tabular}
  \tablefoot{Following stars are removed from the comparisons: $[\mathrm{Fe/H}]_{\rm GALAH}<-1.5$ \& $[\mathrm{Fe/H}]_{\rm This\, study}>-0.5$ in GALAH, and Gaia DR3 4580919415339537280, a relatively hot horizontal branch star being outlier in Figure~\ref{fig:validation}, in \citet{Li2022a}.\\
  \tablefoottext{a}{The median uncertainty in GSP-Spec or our new metallicity estimates. We also provide uncertainties for stars with $[\mathrm{Fe/H}]<-1.5$ in the corresponding metallicities.}\\
  \tablefoottext{b}{The sample selected according to the quality cuts by \citet{RecioBlanco2022a} combined with the cuts on our photometric stellar parameters.}\\
  \tablefoottext{c}{Conditions on GSP-Spec solutions are relaxed from the MP+ sample. More specifically, we remove constraints on the first six digits of the GSP-Spec flags, which are related to vbroad[TGM] and vrad[TGM]. We also include stars with hot $T_{\rm eff, GSP-Spec}$ and high $\log g$ (the region B in Figure~\ref{fig:ew_compare}.}\\
  \tablefoottext{d}{In addition to the RMP sample, this sample includes stars with $KMgiantPar=1$, $[\mathrm{M/H}]_{\rm gspspec}>-3$, and $\log\chi^2_{\rm gspspec}<-2.2$.}\\
  \tablefoottext{e}{In addition to the eRMP1 sample, this sample includes stars with $KMgiantPar=1$, $[\mathrm{M/H}]_{\rm gspspec}=-3.51$.}\\
  }  
  \end{table*}

\begin{figure*}
    \includegraphics[width=1.0\textwidth]{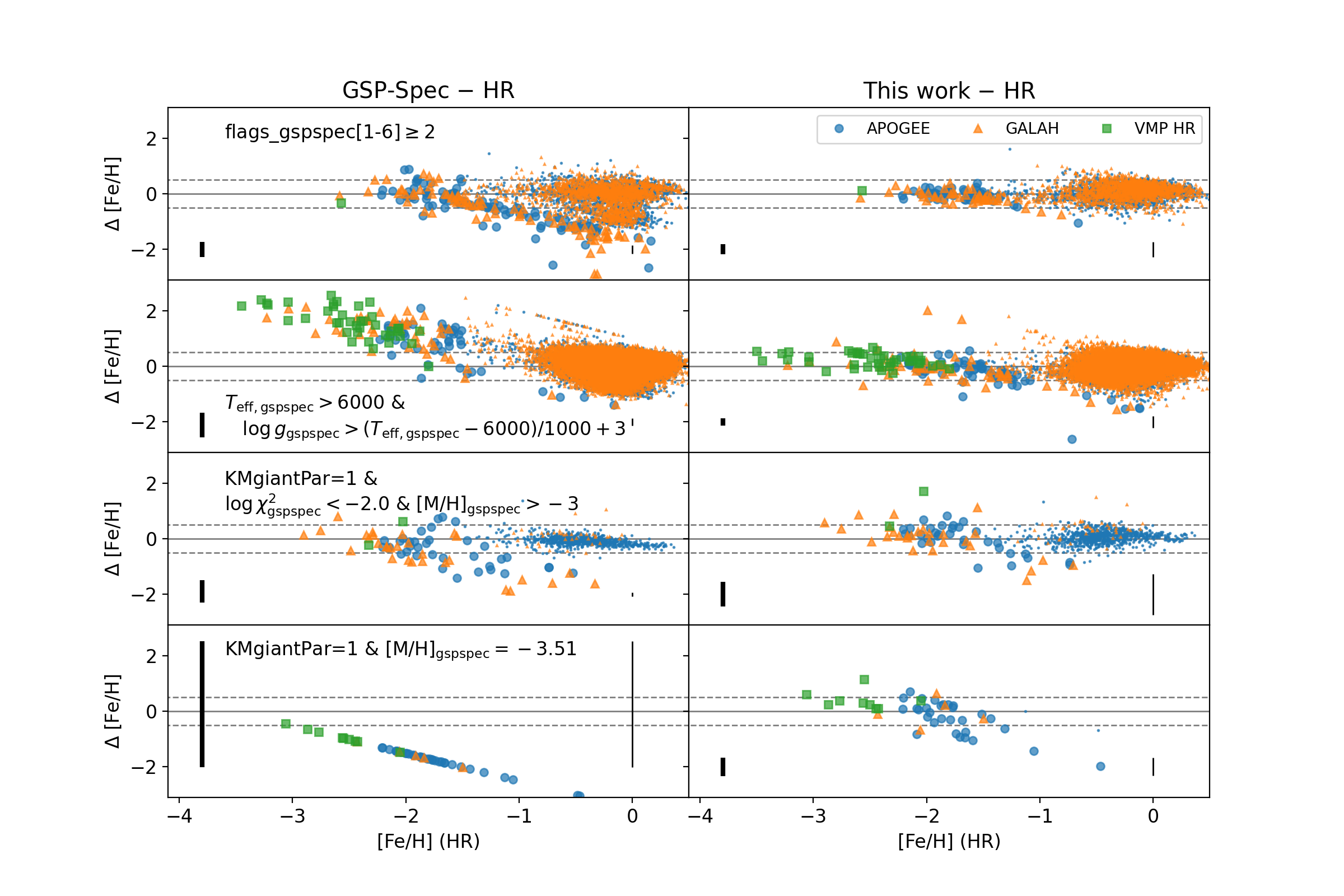}
    \caption{Metallicity comparisons for stars that are added by our new relaxed filters. The text in the figure describes stars newly added in each step. For example, stars in the top panels satisfy all the MP+ filters, but any of \texttt{flags\_gspspec[1-6]} is equal to or larger than 2. We refer the readers to Section~\ref{sec:uncertainty} for more details. We use larger symbol sizes for stars that have $[\mathrm{Fe/H}]<-1.5$ in one of the two metallicities being compared. The grey lines show $\Delta [\mathrm{Fe/H}]=0$ and $\pm 0.5$ dex. We show median metallicity uncertainties for GSP-Spec and our measurements as vertical lines for stars with $[\mathrm{Fe/H}]<-1.5$ on the left and for all the stars on the right. \label{fig:relax}} 
\end{figure*}

In this section, we validate the results via metallicity comparisons with spectroscopic surveys, APOGEE DR17 and GALAH DR3, and the high-resolution study by \citet{Li2022a}.
We remove stars with STAR\_BAD or FE\_H\_FLAG flagged from the APOGEE sample and those with $\texttt{flag\_sp}\neq 0$ or $\texttt{flag\_fe\_h}\neq 0$ from the GALAH sample.
We use the \citet{Li2022a} sample to study our precision at the very metal-poor regime.

The comparisons of the upper panels (GSP-Spec results) and lower panels (our result) of Figure~\ref{fig:validation} show that our method significantly reduces the dispersion at low metallicity and removes outliers.
The lower panels show a bimodal sequence at high metallicity ($[\mathrm{Fe/H}]\gtrsim -0.5$), which can be explained by the $\alpha$-rich and $\alpha$-poor disk populations, and our assumption of the single relation between $[\mathrm{\alpha/Fe}]$ and [{Fe/H}].
\added{The GSP-Spec module provides $[\mathrm{\alpha/Fe}]$ measurements with the typical precision of 0.04 dex.
The measured $[\mathrm{\alpha/Fe}]$ values show a dispersion of 0.11 dex around our assumed $[\mathrm{\alpha/Fe}]$ ratio from Equation \ref{eq:alphafe}, which might be considered as an additional source of error. 
}
Nonetheless, our method still provides reasonable metallicities even for such high-metallicity stars.

One significant improvement is the reduction of cases where the metallicity estimates are far off at low metallicity.
We here investigate the fraction of stars with a metallicity offset larger than $0.5\,\mathrm{dex}$ compared to high-resolution studies without making any quality cuts.
In the published GSP-Spec catalog, the fractions are $76\%$ in the sample of \citet{Li2022a}, and $47\%$ and $42\%$ among the stars with $[\mathrm{Fe/H}]_{\rm HR}<-1.5$ in APOGEE and GALAH.
With our newly derived metallicities, these fractions drop to $24$, $29$, and $19\%$, respectively.

While the improvement at low metallicity is clear, there are stars for which our metallicities still show significant disagreement when compared to high-resolution values, indicating that we need to filter out some stars.
\added{We adopt quality cuts described in the previous section to ensure that our updated stellar parameters do not suffer from reddening, poor photometry, or poor astrometry, that stars' Ca triplet features are sensitive to metallicity and the dominant feature in RVS spectra, and that we can recover the Ca triplet equivalent widths from the published GSP-Spec parameters.}

\added{We first discuss MP+ filters, which are combination of the quality cuts introduced by \citet{RecioBlanco2022a} and those on photometric stellar parameters.
It is clear in Figure~\ref{fig:validation} that the MP+ filters essentially remove stars with a poor agreement.
}
While both the original GSP-Spec metallicity and our new metallicity show good agreements with high-resolution studies within this MP+ sample, there is still a noticeable improvement in our metallicity compared to the GSP-Spec value: we have fewer false VMP stars than the GSP-Spec analysis.
Even with the MP+ filtering, there is a clump of stars at $[\mathrm{M/H}]_{\rm gspspec}\sim -2$ and $[\mathrm{Fe/H}]_{\rm HR}\sim -0.5$ in the top panels of APOGEE and GALAH comparisons, which are false VMP stars.
This clump disappears in the corresponding lower panels.

\added{Provided with the result of the previous section, we here replace the MP filters with more relaxed sets of conditions. 
We remove or relax some of the filters step-by-step and validate each step in Figure~\ref{fig:relax}.
The first two rows of Figure~\ref{fig:relax} correspond to the RMP filters, and the third and last rows correspond to the eRMP1 and eRMP2 filters, respectively.
The figure demonstrates that the agreements remain great for the RMP sample, and reasonably well for eRMP samples.
}

We quantify these agreements with APOGEE, GALAH, and VMP stars from \citet{Li2022a} for the samples in Table~\ref{tab:result}.
\added{It is clear that while our new metallicity agrees with APOGEE and GALAH as good as GSP-Spec original metallicity for the entire MP+ sample, the agreement is significantly improved at low metallicity.
At $[\mathrm{Fe/H}]<-1.5$, our metallicity has much smaller offsets ($-0.01-0.02$ dex compared to $0.15-0.14$ dex) and smaller dispersions ($0.17-0.19$ dex compared to $0.36-0.38$ dex).
In the comparison with \citet{Li2022a}}\footnote{We note that we remove Gaia DR3 4580919415339537280 from the comparison. This star is a hot $T_{\rm eff}\sim 5800\,\mathrm{K}$ horizontal branch star and the outlier in the bottom right panel of Figure~\ref{fig:validation}.}\added{, our metallicity has a larger offset (0.21 dex compared to 0.13 dex) and a smaller dispersion (0.29 dex compared to 0.46 dex). }
This might indicate that there is a metallicity-dependent offset because of systematic uncertainties in the analysis (see Section~\ref{sec:caveats}) since the sample of \citet{Li2022a} is more biased toward low metallicity than the two surveys. 
Alternatively, it is also possible that the metallicity scale of \citet{Li2022a} has an offset from APOGEE and GALAH surveys.

\added{Our estimate of uncertainties seems reasonable for low metallicity stars.
At $\mathrm{Fe/H}]<-1.5$, our new metallicity has the typical uncertainty of 0.18 dex, which is about 55\% of the typical GSP-Spec metallicity uncertainty and close to the dispersions found in the comparisons with APOGEE and GALAH.
Although it is smaller than the dispersion in the comparison with \citet{Li2022a}, this sample has the largest uncertainty among the literature measurements we use (0.11 dex).
Therefore, we consider that our estimates of uncertainties are reasonable and that our method provides more precise metallicities for low-metallicity stars than the GSP-Spec module.
}

\added{
On the other hand, the typical metallicity uncertainty is much larger in our method (0.32 dex) than in GSP-Spec (0.06 dex) for the entire sample.
This indicates that the GSP-Spec module performs excellently for high metallicity stars because there is enough information in their RVS spectra and it already utilizes spectral features to constrain stellar parameters.
Despite the larger uncertainty in our method, the dispersions in comparisons remain small, indicating that our uncertainty is likely overestimated at high metallicity.
The overestimation is not unexpected since we used stars with $[\mathrm{Fe/H}]<-2$ to estimate uncertainties in the recovered equivalent widths and the recovered equivalent widths might have higher accuracy at higher metallicity.
Since the GSP-Spec already provides stellar parameters with high quality for high metallicity stars, one can adopt GSP-Spec metallicities for such stars, and hence we do not try to provide more realistic uncertainties for high metallicity stars.
The similar dispersions in the two comparisons are also expected since our method does not return a significantly different metallicity if GSP-Spec stellar parameters already agree with photometric estimates, which would likely be the case for high metallicity stars that do not have degeneracy issues.
}

Table~\ref{tab:result} also contains comparisons with RMP and eRMP samples.
It is clear that the GSP-Spec original metallicities show a large scatter at $[\mathrm{Fe/H}]<-1.5$ when we use the RMP and eRMP samples, while our new metallicity estimates in the three samples are as good as in the MP+ sample. 
We also note that the numbers of VMP stars from \citet{Li2022a} in RMP and eRMP samples increase significantly compared to the MP sample (from 22 to 64 and 76 stars used for the comparisons, respectively).
This demonstrates the power of photometric information for the analysis of low-metallicity stars' spectra.

We finally comment on stars that have high metallicity in our estimates ($[\mathrm{Fe/H}]>-0.5$) but low metallicity in GALAH ($[\mathrm{Fe/H}]<-1.5$).
Most of these stars are either around $(\alpha,\,\delta)=(155.63^\circ,\,-44.31^\circ)$ with the GALAH \textit{field\_id} 4205 or around $(225.93^\circ,\,-77.02^\circ)$ with the \textit{field\_id} 2620. 
\added{Since they have 2 in the 11th digit of \textit{sobject\_id}, they might have been affected by the stacking issue in the GALAH DR3 \citep{Buder2021a}.}

\section{Discussion}\label{sec:discussion}
\subsection{Properties of our VMP stars}
\begin{figure}
\includegraphics[width=0.45\textwidth]{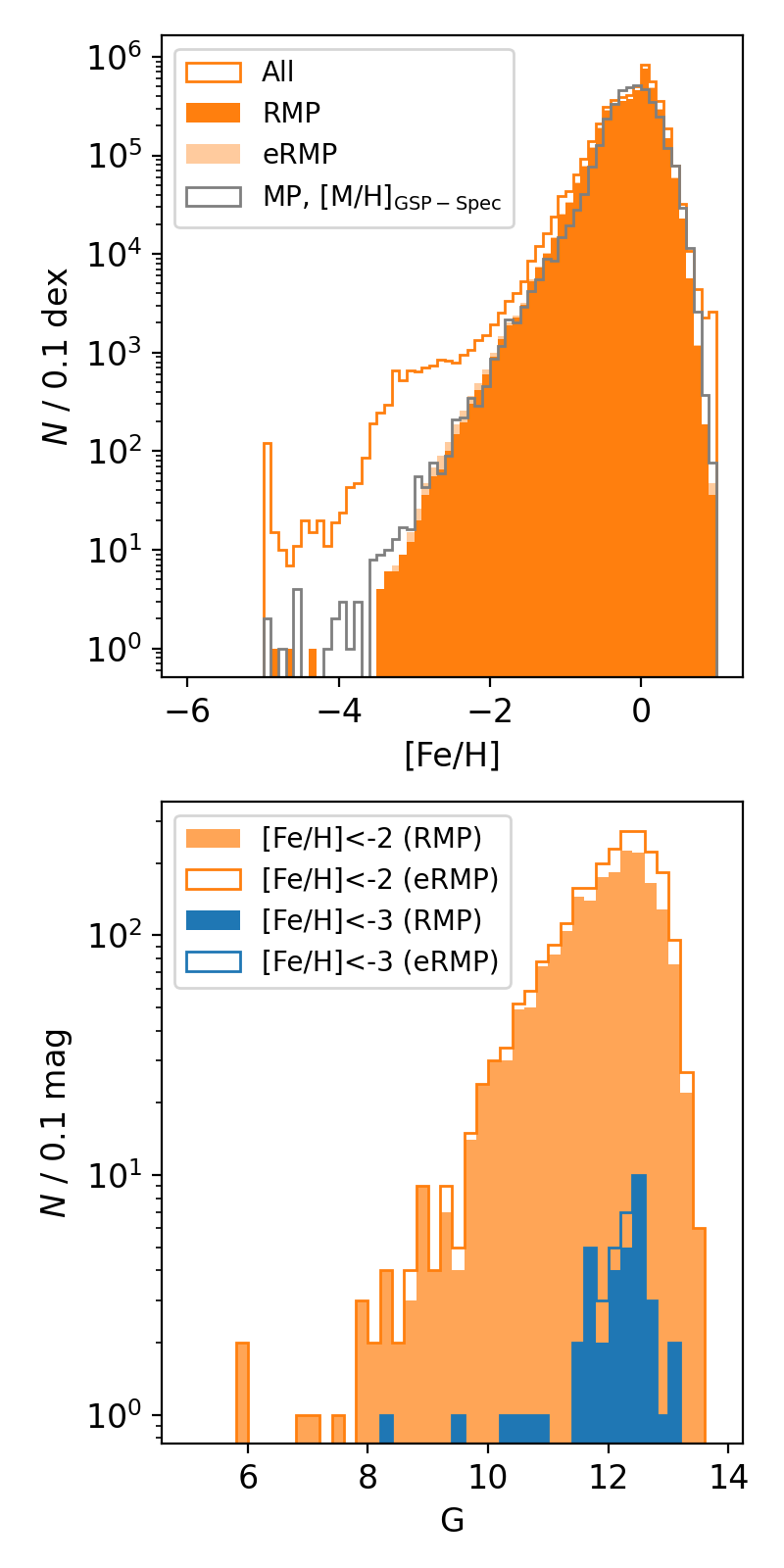}
\caption{Metallicity and magnitude distributions of stars. (Top:) Orange histograms show the metallicity distribution functions in our new metallicities with different sets of filters. The grey line indicates the MDF from the  GSP-Spec [{M/H}]. (Bottom:) Magnitude distribution of our VMP and EMP stars in RMP and eRMP filters.\label{fig:MDFmag}}
\end{figure}

In this section, we summarise the properties of our VMP stars in RMP and eRMP2 samples and compare them with those of VMP stars in the MP sample.
We focus on the metallicity and magnitude distributions and how the catalog of VMP stars changes by moving from the GSP-Spec analysis to our analysis.

Figure~\ref{fig:MDFmag} shows the metallicity distribution functions for our RMP and eRMP2 samples created with our new metallicity estimates, which are also compared with the original $[\mathrm{M/H}]_{\rm gspspec}$ distribution from the MP sample. 
The number of VMP stars and extremely metal-poor stars (EMP stars with $[\mathrm{Fe/H}]<-3$) remain more or less the same, and the overall shape of metallicity distribution is unchanged.
We find 1941 VMP and 90 EMP stars with $[\mathrm{M/H}]_{\rm gspspec}$ and the MP-filters, 1989 VMP and 40 EMP stars in our new metallicities with RMP filters, and 2345 VMP and 43 EMP stars with eRMP filters.
Although the number of VMP stars remains similar, only 895 in the first sample of VMP stars remain in the RMP VMP sample.
As noted in Section~\ref{sec:feh_comparison}, this is partly thanks to the removal of false VMP stars when moving from the GSP-Spec metallicity to our measurements.
The remaining $\sim 1000$ stars are now not regarded as VMP stars in our new metallicity estimates.
The reduction in the number of VMP stars is then compensated by our relaxed filters described in Section~\ref{sec:uncertainty}.

The relaxed filters are particularly efficient in keeping known low metallicity stars.
In addition to the significant increase in the overlap with the \citet{Li2022a} sample mentioned above, we can also confirm the effect of the relaxed filters by looking at how many VMP and EMP stars in RMP and eRMP2 samples are in the MP sample.
Of the 1989 VMP and 40 EMP stars in the RMP sample, only 71\% and 23\%  are in the MP sample. 
The corresponding numbers for the eRMP sample are 60\% and 21\%.
About $30-40\%$ of VMP stars and $\sim 80\%$ of EMP stars could not have been identified if we did not try to relax the MP filters.

Our RMP sample offers one of the best and brightest metal-poor star catalogs to date.
As they have been analyzed by the GSP-Spec module, most stars are brighter than $G=13$ (Figure~\ref{fig:MDFmag}).
Such bright stars are easily followed up with high-resolution spectroscopy for detailed abundance measurements.
One can also use the eRMP samples, but \added{they have larger uncertainties}.

\subsection{Caveat and future prospects}\label{sec:caveats}

Although we have significantly improved the accuracy of metallicity for very metal-poor stars, there are limitations in this approach.
First, the spectra are unavailable for $\sim 80\%$ of the stars, making it impossible to start directly from spectra or visually check the fitting results.
While there is a statistically good agreement between high-resolution and our new metallicities, there could be very rare cases where the spectrum and/or the stellar parameters have issues, for example, because of flux contamination from nearby bright stars, presence of unresolved companions, and/or stellar activity and rotation.
Visually checking the fitting result is useful, especially when one wants to remove such rare cases for follow-up observations of high-confidence EMP stars. 
The fraction of such stars is to be estimated from such follow-up observations.

Second, while the use of photometric information greatly contributes to breaking the degeneracy, it introduces a dependency on photometric measurements and extinction estimates.
One of the advantages of fully spectroscopic analysis is that one can analyze all the spectra consistently without being affected by the coverage of photometric surveys and dust maps. 
While our approach basically cancels this advantage of the GSP-Spec module, we consider that the gain more than compensates it for very metal-poor stars.

While we use external information to Gaia for the photometric information, it should, in principle, be possible to obtain photometric stellar parameters from the Gaia BP/RP spectra.
We refrain from using the GSP-Phot analysis results of the Gaia BP/RP spectra since they currently suffer from systematic offsets \citep{Andrae2022a}.
Thus, future improvements in the GSP-Phot module would also benefit the analysis of RVS spectra of VMP stars.

We here provide new metallicities based on the local thermodynamic equilibrium (LTE) assumption.
Since non-LTE effects are known to be significant for the Ca triplet \citep{Mashonkina2007a,Sitnova2019a,Osorio2022a}, our metallicities are subject to deviation from the LTE assumption.
We here stick to the LTE analysis since grids of the correction do not span the whole parameter space we explore, and we aim to keep consistency with the GSP-Spec analysis as much as possible, which adopts the LTE assumption.
Since the abundance obtained from LTE analyses of the Ca triplet is generally higher than non-LTE analyses, our metallicities are likely overestimated. 
The metallicity dependence of this effect may be related to the median offset we find only in comparison with \citet{Li2022a} in Table~\ref{tab:result}.

One challenge we have is how to include more stars with $KMgiantPar$ flagged. 
Many VMP stars are mistaken with higher metallicity cool giants with significant molecule features.
The latter type of stars mimics the spectra of VMP stars with weak and shallow absorption lines due to a pseudo continuum.
A better understanding of how we can separate these types of stars would significantly benefit the search for VMP and EMP stars from the RVS spectra.

Since we no longer need to break the degeneracy between stellar parameters, the use of photometric information may allow us to derive metallicities from RVS spectra with a lower signal-to-noise ratio than the current threshold.
As we see in the sample size of stars with radial velocity measurements in Gaia DR2 \citep[down to $G\sim 13$, 7.2 million stars][]{GaiaCollaboration2018a} and Gaia DR3 \citep[down to $G\sim 14$, 33 million stars][]{GaiaCollaboration2022a}, we expect a significant increase in the sample size by going deeper.

\section{Summary}\label{sec:summary}

While the GSP-Spec module provides precise spectroscopic stellar parameters and chemical abundances for most of the sample, the analysis suffers \added{from parameter degeneracy} at the very low metallicity regime.
\added{As a result, many of known very metal-poor stars have large metallicity uncertainty and fail to pass quality cuts.}
We mitigate this difficulty by incorporating the photometric information into the analysis\added{, provide improved metallicities with smaller uncertainty for low-metallicity stars, and aim to keep more very metal-poor stars even after some quality cuts.}
We first compute the Ca triplet equivalent widths from the published set of GSP-Spec parameters using Turbospectrum.
We then convert the equivalent widths back to metallicity assuming photo-astrometric $T_{\rm eff}$ and $\log g$.
Our new Ca triplet metallicity shows a better agreement with the high-resolution value for the sample of very metal-poor stars by \citet{Li2022a}.

Our main findings can be summarized as follows:
\begin{itemize}
\item We conduct an independent validation of the GSP-Spec metallicities using the high-resolution spectroscopic study of very metal-poor stars by \citet{Li2022a}. For a sample satisfying the recommended set of filters by \citet{RecioBlanco2022a}, the GSP-Spec metallicity shows a reasonable agreement with \citet{Li2022a}. However, this set of filters (MP filters) leaves only 23 stars out of 109 stars with high-resolution and GSP-Spec metallicities. \added{Moreover, even the 23 stars have large uncertainty (median uncertainty of 0.44 dex). Despite even larger uncertainties (0.82 dex),} stars that do not satisfy this quality cut \added{show metallicity and temperature disagreement beyond the measurement uncertainties, which would likely be due to} parameter degeneracy (Figure~\ref{fig:gspspec_VMP}). 
\item Even if a star suffers from parameter degeneracy, we can reproduce the strengths of the Ca triplet reasonably well with the published GSP-Spec parameters (Figure~\ref{fig:spec} \added{and Section~\ref{sec:ew_validation}}). \added{We confirmed this by directly measuring the Ca triplet equivalent widths from public Gaia RVS spectra and comparing them with our reproduced values. We further investigated how the precision of the recovered equivalent widths changes in GSP-Spec stellar parameter space and defined subsamples for which we can reasonably recover the equivalent widths (Section~\ref{sec:uncertainty}). 
These subsamples include stars that are recommended to filter out by \citet{RecioBlanco2022a}, such as stars with $T_{\rm eff, GSP-Spec}>6000\,\mathrm{K}$ and cool K and M-type giant stars.}
\item This motivates us to 
re-derive metallicities \added{from the recovered Ca triplet equivalent widths} adopting a new set of $T_{\rm eff}$ and $\log g$ that are derived from photometry and astrometry.
\item The new Ca triplet metallicity shows a good agreement with high-resolution values at low metallicity and a reasonable agreement even at high metallicity (Figure~\ref{fig:validation}). 
\added{The new metallicity has a typical uncertainty of 0.18 dex at $[\mathrm{Fe/H}]<-1.5$, which is 45\% smaller than the median GSP-Spec uncertainty at the same metallicity.}
It also reduces the fraction of catastrophic failure at low metallicity, where the metallicity estimates are different from high-resolution values by more than $0.5\,\mathrm{dex}$.
This fraction is now $24\%$ with our new metallicity, while it is $76\%$ in the original GSP-Spec metallicity.

\item 
We are now able to consider our metallicity to be reliable for 64 stars out of the aforementioned 109 very metal-poor stars. 
This number goes up to 76 if we adopt our most extended sample. 
For the sample of 64 and 76 stars, the mean differences ($\pm$ standard deviation) compared to the \citet{Li2022a} values are $0.22$ ($\pm 0.25$) and $0.24$ ($\pm 0.34$) dex, respectively. 
\item Our sample offers one of the best catalogs of very bright low-metallicity stars. The numbers of VMP and EMP stars are 1989 and 40 stars. In the most extended sample, these numbers increase to 2345 and 43, but it comes with a low level of metal-rich contaminants.
We make the catalog available online. 
\end{itemize}

As discussed in Section~\ref{sec:caveats}, while the use of photometric information comes with merits, especially for stars with little information in their spectra due to low metallicity or low signal-to-noise ratio, it introduces a dependency on external data sets.
For example, the present work relies on 2MASS and extinction maps from multiple studies.
Future data releases from Gaia with improved GSP-Photo stellar parameters might enable us to conduct a similar analysis to the present study: a spectroscopic analysis with photometric information only using Gaia data. 

\begin{acknowledgements}

We thank Georges Kordopatis at Universit\'{e} C\^ote d'Azur for his helpful comments.
This research has been supported by a Spinoza Grant from the Dutch Research Council (NWO).
ES acknowledges funding through VIDI grant ``Pushing Galactic Archaeology to its limits'' (with project number VI.Vidi.193.093) which is funded by the Dutch Research Council (NWO).
We benefited from support from the International Space Science Institute (ISSI) in Bern, CH, thanks to the funding of the Team ``the early Milky Way''.

This work has made use of data from the European Space Agency (ESA) mission
{\it Gaia} (\url{https://www.cosmos.esa.int/gaia}), processed by the {\it Gaia}
Data Processing and Analysis Consortium (DPAC,
\url{https://www.cosmos.esa.int/web/gaia/dpac/consortium}). Funding for the DPAC
has been provided by national institutions, in particular the institutions
participating in the {\it Gaia} Multilateral Agreement.

\end{acknowledgements}

\begin{appendix}

\section{Summary of our quality cuts}\label{appendix:qc}
In the main text, we define the following samples.

\noindent MP+ sample, which is combination of the filters by \citet{RecioBlanco2022a} and filtering on photometric parameters:
\begin{itemize}
\item $3500\leq T_{\rm eff, gspspec}\leq 6000$
\item $\log g_{\rm gspspec}\leq 4.9$
\item $T_{\rm eff, gspspec}\geq 4150$ or $\log g_{\rm gspspec}\leq 2.4$ or $\log g_{\rm gspspec}\geq 3.8$
\item vbroad[TGM]=0 and vrad[TGM]=0
\item $extrapol\leq 2$
\item $KMgiantPar=0$ 
\item \texttt{spectraltype\_esphs} is not O or B
\item $2d_{\rm photogeo,med}/(d_{\rm photogeo,hi}-d_{\rm photgeo,lo})>3$
\item $E(B-V)<0.72$
\item $T_{\rm eff, photo}<7000$
\item $T_{\rm eff, photo}>5000$ or $\log g_{\rm photo}<3.0$.
\end{itemize}

\noindent RMP sample, which relaxes or removes some of the conditions in the MP+ filtering:
\begin{itemize}
  \item $3500\leq T_{\rm eff, gspspec}$
  \item $T_{\rm eff, gspspec}\leq 6000$ or $\log g_{\rm gspspec}\geq (T_{\rm eff,gspspec}-6000)/1000 +3$
  \item $T_{\rm eff, gspspec}\geq 6000$ or $\log g_{\rm gspspec}\leq 4.9$
  \item $T_{\rm eff, gspspec}\geq 4150$ or $\log g_{\rm gspspec}\leq 2.4$ or $\log g_{\rm gspspec}\geq 3.8$
  \item $extrapol\leq 2$
  \item $KMgiantPar=0$   
  \item \texttt{spectraltype\_esphs} is not O or B
  \item $2d_{\rm photogeo,med}/(d_{\rm photogeo,hi}-d_{\rm photgeo,lo})>3$
  \item $E(B-V)<0.72$
  \item $T_{\rm eff, photo}<7000$
  \item $T_{\rm eff, photo}>5000$ or $\log g_{\rm photo}<3.0$.
\end{itemize} 

\noindent eRMP1 sample, which extends the RMP sample by including some stars with KMgiantPar flagged:
\begin{itemize}
  \item $3500\leq T_{\rm eff, gspspec}$
  \item $T_{\rm eff, gspspec}\leq 6000$ or $\log g_{\rm gspspec}\geq (T_{\rm eff,gspspec}-6000)/1000 +3$
  \item $T_{\rm eff, gspspec}\geq 6000$ or $\log g_{\rm gspspec}\leq 4.9$
  \item $T_{\rm eff, gspspec}\geq 4150$ or $\log g_{\rm gspspec}\leq 2.4$ or $\log g_{\rm gspspec}\geq 3.8$
  \item $extrapol\leq 2$
  \item $KMgiantPar=0$ or ($KMgiantPar=1$ and $\log \chi^2_{\rm gspspec}<-2.0$ and $[\mathrm{M/H}]_{\rm gspspec}>-3$)
  \item \texttt{spectraltype\_esphs} is not O or B
  \item $2d_{\rm photogeo,med}/(d_{\rm photogeo,hi}-d_{\rm photgeo,lo})>3$
  \item $E(B-V)<0.72$
  \item $T_{\rm eff, photo}<7000$
  \item $T_{\rm eff, photo}>5000$ or $\log g_{\rm photo}<3.0$.
\end{itemize} 

\noindent eRMP2 sample, which further extends the eRMP1 sample by including more stars with KMgiantPar flagged:
\begin{itemize}
  \item $3500\leq T_{\rm eff, gspspec}$
  \item $T_{\rm eff, gspspec}\leq 6000$ or $\log g_{\rm gspspec}\geq (T_{\rm eff,gspspec}-6000)/1000 +3$
  \item $T_{\rm eff, gspspec}\geq 6000$ or $\log g_{\rm gspspec}\leq 4.9$
  \item $T_{\rm eff, gspspec}\geq 4150$ or $\log g_{\rm gspspec}\leq 2.4$ or $\log g_{\rm gspspec}\geq 3.8$
  \item $extrapol\leq 2$
  \item $KMgiantPar=0$ or ($KMgiantPar=1$ and $\log \chi^2_{\rm gspspec}<-2.0$ and $[\mathrm{M/H}]_{\rm gspspec}>-3$) or ($KMgiantPar=1$ and $[\mathrm{M/H}]_{\rm gspspec}=-3.51$)
  \item \texttt{spectraltype\_esphs} is not O or B
  \item $2d_{\rm photogeo,med}/(d_{\rm photogeo,hi}-d_{\rm photgeo,lo})>3$
  \item $E(B-V)<0.72$
  \item $T_{\rm eff, photo}<7000$
  \item $T_{\rm eff, photo}>5000$ or $\log g_{\rm photo}<3.0$.
\end{itemize} 

\section{Curve of growth for Ca 8552 and 8662 lines.}\label{appendix:cog}

\begin{figure*}
\centering
\includegraphics[width=\linewidth]{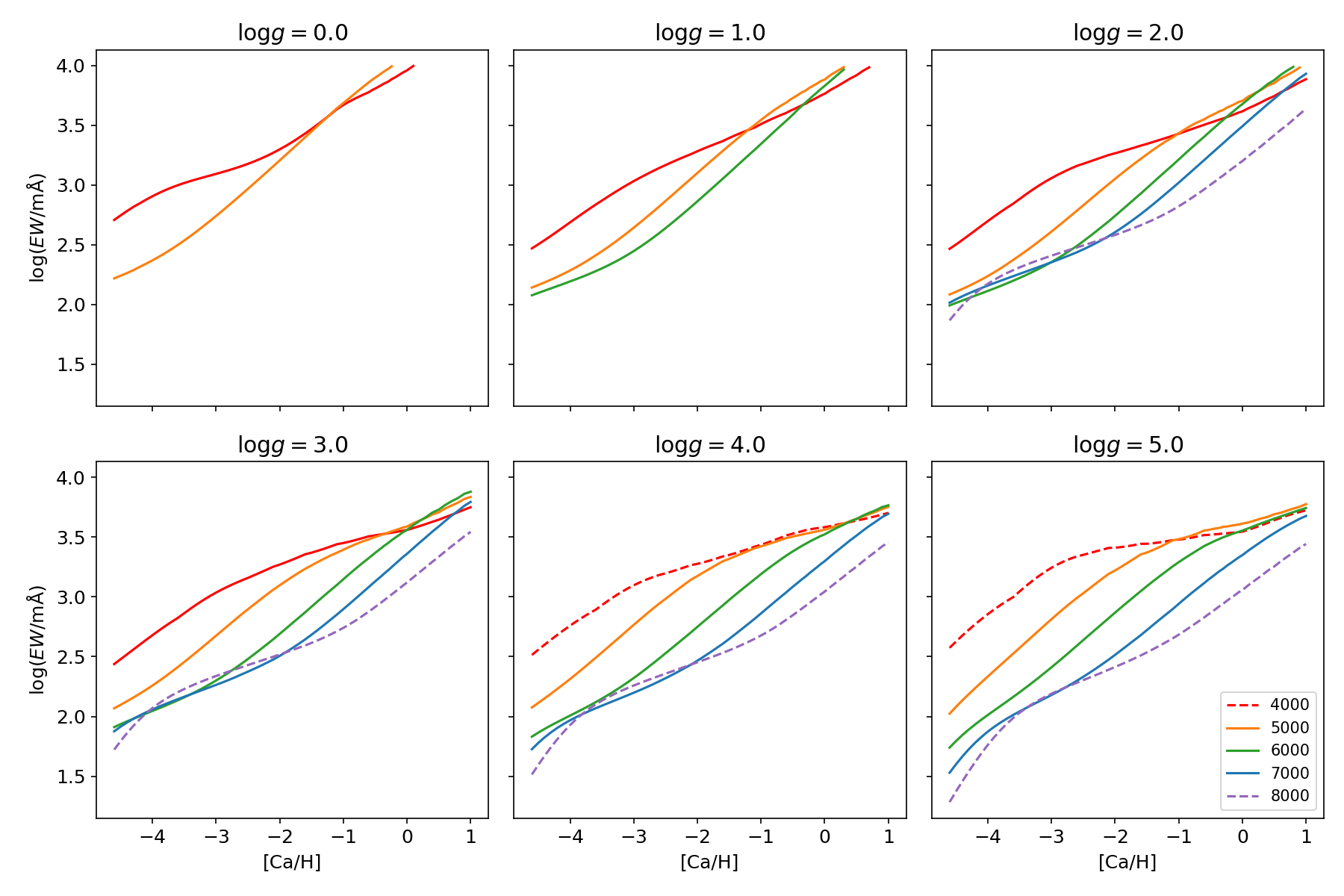}
\caption{Same as Figure~\ref{fig:sensitivity} but for Ca $8542\,\mathrm{\AA}$.}
\end{figure*}
\begin{figure*}
\centering
\includegraphics[width=\linewidth]{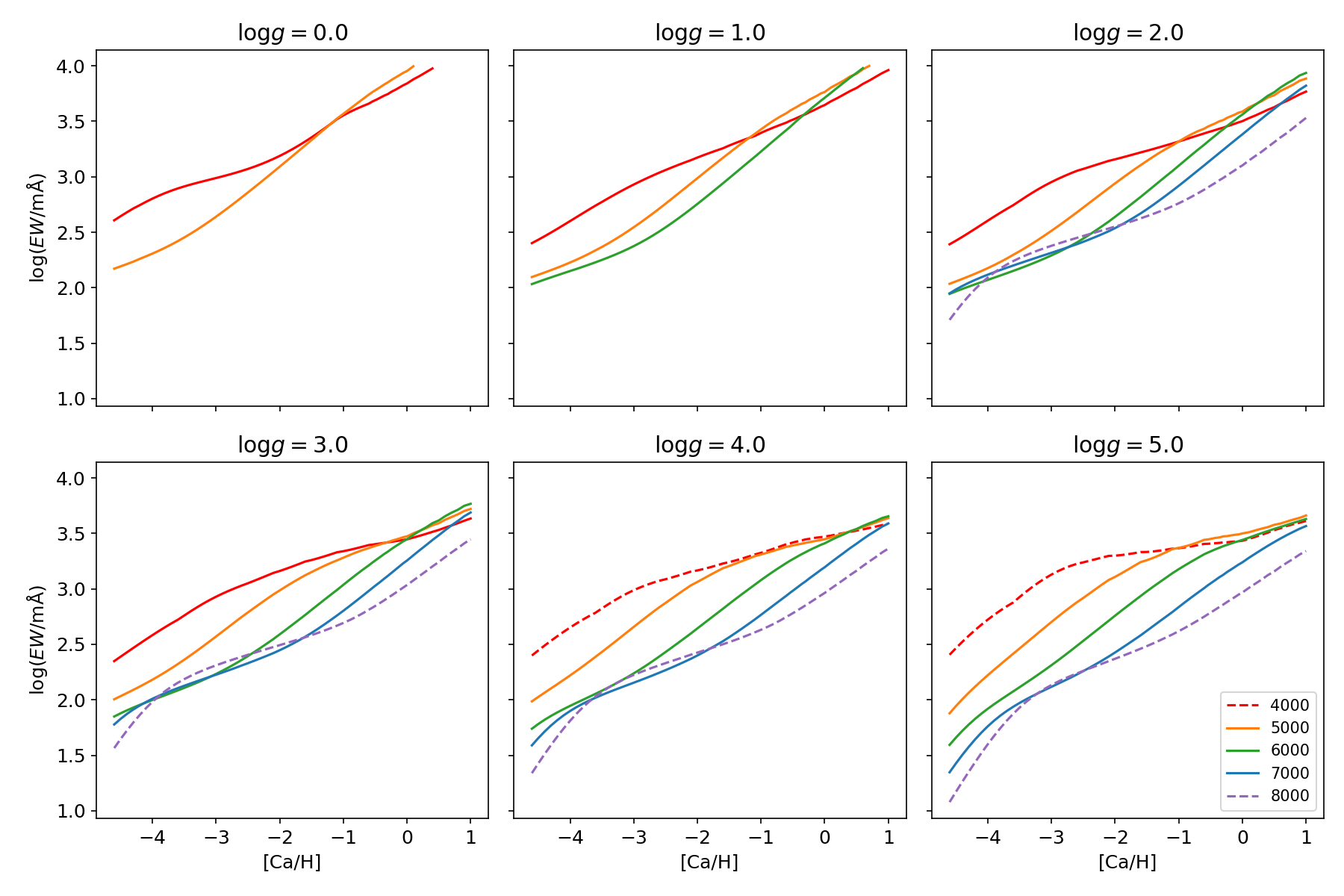}
\caption{Same as Figure~\ref{fig:sensitivity} but for Ca $8662\,\mathrm{\AA}$.}
\end{figure*}

\end{appendix}
\end{document}